\documentclass{aa}  

\usepackage{graphicx}
\usepackage{txfonts}
\usepackage{xcolor}
\usepackage{lastpage}

\usepackage{amsmath}
\usepackage[utf8]{inputenc}
\usepackage{stfloats}

\DeclareRobustCommand{\pz}{\emph{pz}}
\DeclareRobustCommand{\sz}{\emph{sz}}
\DeclareRobustCommand{\msun}{\mathrm{M_\odot}}

\defcitealias{shen_size_2003}{S03}
\defcitealias{ghosh_denser_2024}{G24}
\defcitealias{wen_catalog_2024}{WH2024}

\begin{document} 

   \title{Environmental Invariance of the Galaxy Size–Mass Relation}


   \author{Li-Wen Liao
          \inst{1}\fnmsep\inst{2}
          \and
          Andrew P. Cooper \inst{3,4}
          }

   \institute{Institute of Space Sciences (ICE, CSIC), Campus UAB, Carrer de Can Magrans, s/n, 08193 Barcelona, Spain\\
   \email{lwliao@ice.csic.es}
   \and
   Institut d’Estudis Espacials de Catalunya (IEEC), Edifici RDIT, Campus UPC, 08860 Castelldefels (Barcelona), Spain
    \and
    Institute of Astronomy and Department of Physics, National Tsing Hua University, Hsinchu 30013, Taiwan\\
    \email{apcooper@gapp.nthu.edu.tw}
    \and
    Center for Informatics and Computation in Astronomy, National Tsing Hua University, Hsinchu 30013, Taiwan 
             }

   \date{Received August 05, 2025; accepted November 17, 20xx}

 
  \abstract
   {
   The galaxy size-luminosity and size-stellar mass relations are important constraints on the galactic baryon cycle of gas accretion, star formation, and feedback.
   There are conflicting claims in the literature regarding how environment influences size: both `direct' transformative effects and `assembly bias' may contribute to observed variations with environment.
   }
   {
   We construct a large homogeneous sample of size measurements to $M_r \sim -14$ ($M_\star \sim 10^{7}$ $\msun$).
    Our sample fills a gap in field galaxy size measurements around $\sim10^{7}$ -- $10^{8}\,\mathrm{\msun}$; the literature at these masses is biased towards satellites of $L^\star$ galaxies and members of galaxy clusters.
   }
   {
   We use sizes from the DESI Legacy Survey (DESI-LS; significantly larger and deeper than SDSS), together with a published catalog that contains stellar masses and cluster positions derived from DESI-LS photometry. Our sample extends to $z < 0.3$ and comprises 540,228 galaxies with spectroscopic redshifts and 9,513,732 galaxies with photometric redshifts. We explore the environmental dependence of size for a mass-limited subset of our sample at $z < 0.05$, based on distance to the nearest cluster center.
   }
   {
   We obtain size-luminosity and size-mass relations in good agreement with previous studies. By separating galaxies according to color and morphology, we show that the environmental variation of the overall size-mass relation on megaparsec scales can be understood as the consequence of a changing mixture of subpopulations, rather than `direct' size transformation. For example, at fixed mass, quiescent (red) late-type galaxies within $2$~Mpc of a cluster have the same size as quiescent late-type galaxies $30$~Mpc from the nearest cluster. 
   }
   {Our results support individual galaxy assembly histories as the primary determinant of galaxy size. The existence of significantly different, environment-insensitive size mass relations for subpopulations separated by color (star formation rate) and S\'ersic index (morphology) provides a clear target for calibration of the baryon cycle in cosmological simulations.}

   \maketitle
%

\section{Introduction}

\begin{figure*}
    \sidecaption
    \includegraphics[width=12cm]{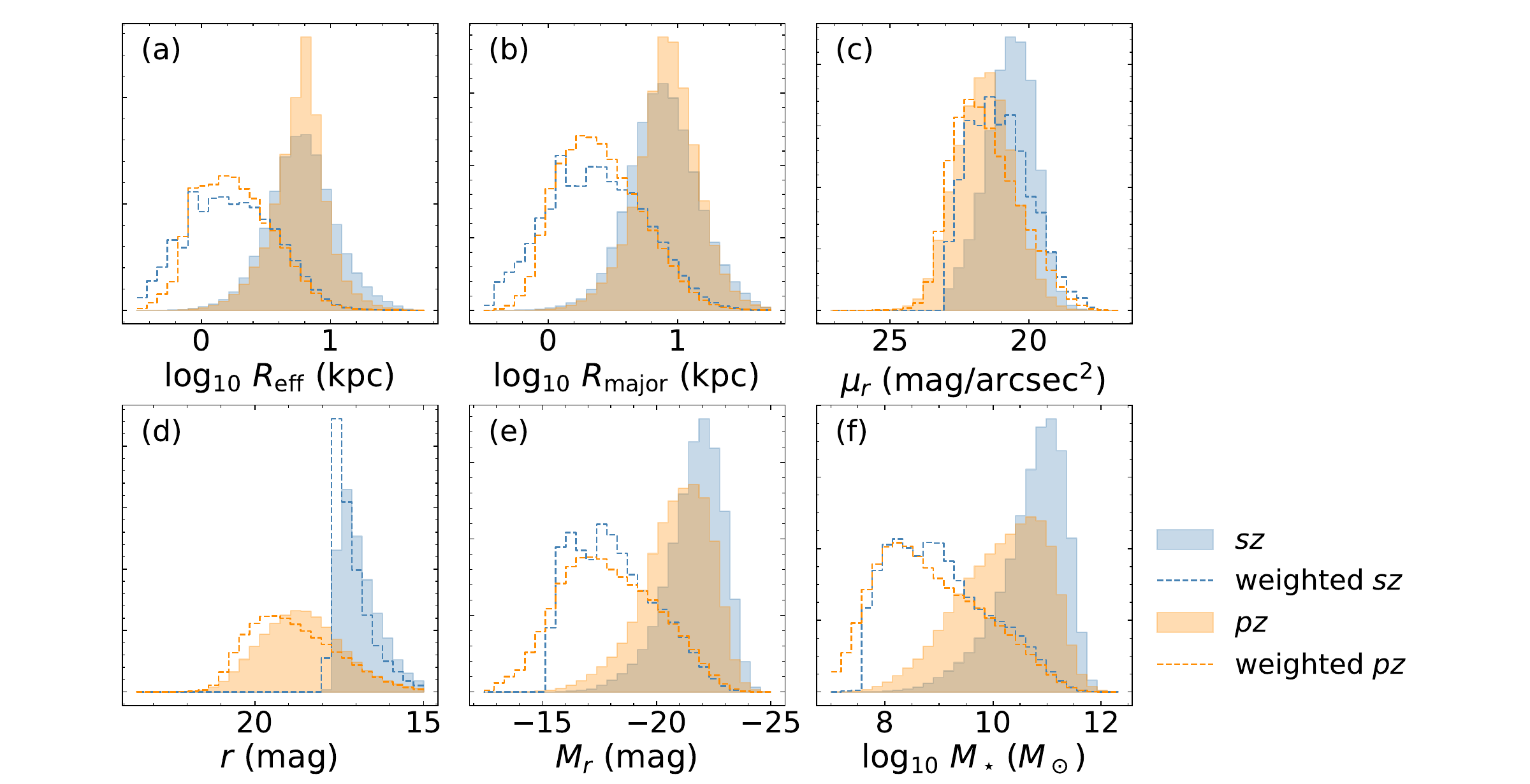}
    \caption{A summary of the two main galaxy samples we use in this work. Normalized histograms of (a) $R_\mathrm{eff}$, (b) $R_\mathrm{maj}$, (c) $\mu_r$, (d) $r$-band magnitude, (e) $M_r$, and (f) $M_\star$ for data with spectroscopic redshift and data with photometric redshift. Filled blue and filled orange show the unweighted distribution of data with spectroscopic redshift and data with photometric redshift. The dashed blue and orange histograms represent the $V/V_\mathrm{max}$ weighted distributions of the spectroscopic and photometric data samples, respectively. The density scale on the vertical axis is arbitrary. }
    \label{fig:data_dis}
\end{figure*}

The relationships between galaxy properties, such as size, luminosity, and morphology, provide fundamental insights into galaxy formation and evolution \citep[see, for example, reviews by][]{blanton_physical_2009, simon_faintest_2019}.
The size-luminosity and size-mass relations are among the most extensively studied relationships for both disk and elliptical galaxies \citep{shen_size_2003, nair_environmental_2010, huertas-company_evolution_2013, bernardi_systematic_2014, van_der_wel_3d-hstcandels_2014, lange_galaxy_2016, chen_sizemass_2022, yoon_low-mass_2023}.
Large statistical studies have been used to obtain size-mass relations for
galaxies with stellar masses greater than $10^9\, \mathrm{M_\odot}$ \citep{mosleh_stellar_2018, mowla_cosmos-dash_2019, hon_size-mass_2023, cook_deep_2025}.
However, most studies of the sizes of intrinsically faint and low-surface brightness galaxies have been restricted to the Local Group \citep{brasseur_what_2011, mcconnachie_observed_2012, doliva-dolinsky_pandas_2023} and in nearby clusters, such as Virgo \citep{toloba_formation_2012, jang_discovery_2014, mihos_galaxies_2015}, Fornax \citep{munoz_unveiling_2015}, and Coma \citep{van_dokkum_forty-seven_2015}.
Consequently, at the extremely faint end \citep[as faint as $M_V \sim 0$;][]{mcconnachie_observed_2012}, the available data on galaxy sizes comprise a variety of heterogeneous samples from different surveys and environments.
There is a substantial gap at intermediate luminosities ($M_V\sim-15$) between these dwarf galaxy data and the size-mass relations from large surveys of brighter galaxies
\citep{brasseur_what_2011}.

Larger, deeper wide-field surveys will bridge this gap.
For example, the wide-field component of the Hyper Suprime-Cam Subaru Strategic Program Survey (HSC-SSP) covers 1,200 square degrees to a point-source detection limit of $r \sim 26$ \citep{aihara_third_2022} with an effective surface brightness limit of $\mu\sim$ 30 mag arcsec$^{-2}$ \citep{thuruthipilly_hsc_2025}, and the DESI Legacy Imaging Survey (LS) reaches a point-source limit of $r \sim 23.9$ over $\sim$20,000 square degrees \citep{dey_overview_2019}, with an effective surface brightness limit of $\mu\sim29$ mag arcsec$^{-2}$ \citep{tanoglidis_shadows_2021}. In the near future, the space telescope Euclid is expected to provide photometric observations over 14,000 square degrees to a depth of $r \sim 24.5$ and $\mu\sim 29.5$ mag arcsec$^{-2}$ along with spectroscopic data for millions of galaxies \citep{borlaff_euclid_2022,
euclid_collaboration_euclid_2024}. Meanwhile, the ground-based Vera Rubin telescope will survey 18,000 square degrees, reaching a depth of $r \sim 27.5$ and $\mu\sim30.6$ mag arcsec$^{-2}$ \citep{laine_lsst_2018, ivezic_lsst_2019}.

In this paper, motivated by the rapidly growing interest in statistical studies of field dwarf galaxies driven by these upcoming surveys, we take advantage of the DESI Legacy Imaging Survey to measure the size-mass relation for the `field' galaxy population to a magnitude limit fainter than the previous state of the art (the Sloan Digital Sky Survey; SDSS).

The formation of massive galaxies is now well understood, at least in outline, but this is not yet the case for dwarf galaxies.\footnote{For concreteness, we take dwarf galaxies to be those with $M_{V}\gtrsim - 18$; this is quite arbitrary but sufficient for our discussion.}
For example, it is generally agreed that the sizes of disk galaxies are related to their angular momentum \citep{fall_formation_1980, mo_formation_1998}, and the sizes of elliptical galaxies can be understood as the consequence of the virial theorem applied to hierarchical merger events, perhaps influenced also by merger-driven starbursts
\citep{cole_hierarchical_2000, daddi_passively_2005, naab_properties_2006, bernardi_curvature_2011}.
Dwarf galaxies, in contrast, develop irregular morphologies; this is thought to be due to their short cooling time-scales, the high efficiency of supernova-driven gas expulsion in shallow gravitational potentials, and circular speeds comparable to the random velocities in the dense star-forming gas \citep[][]{mo_formation_1998, burger_kinematic_2022, rowland_pre-supernova_2024}. In this low-mass regime, it is unclear how the different components of the `baryon cycle' interact to determine the present-day size of a galaxy. 
Moreover, dwarf galaxies exhibit morphological diversity, for example, the dichotomy between dwarf spheroidals and dwarf irregulars \citep{mateo_dwarf_1998, lazar2024dwarfmorp}. The physical processes that give rise to this diversity remain poorly understood \citep[e.g.,][]{van_nest_whats_2022, zaritsky_photometric_2023, wang_ghostly_2023}.

Some of the diversity of the dwarf galaxy population may be explained by environmental effects, such as tidal stripping, once they become satellites \citep{yaryura_environmental_2023, chamba_impact_2024}. However, the lack of large samples, particularly in the field, limits statistical tests of models in this regime. For example, many size-mass relation studies have considered evidence for environmental effects \citep[e.g.][]{nair_environmental_2010, cebrian_effect_2014, mosleh_stellar_2018, yoon_low-mass_2023, chen_sizemass_2022, chamba_impact_2024, perez_galaxy_2025}.
These effects, in particular tidal stripping in clusters, are expected to be more significant for low-mass galaxies \citep{elmegreen_outskirts_2017}.
However, as yet there is no clear agreement on how or why galaxy sizes may vary systematically with environment \citep[see e.g.\ the discussion in][]{ghosh_denser_2024}. 
Some studies find significant size differences in denser environments
\citep{poggianti_superdense_2013, cebrian_effect_2014, chamba_impact_2024}, while others do not, even at low mass \citep{huertas-company_no_2013, kuchner_effects_2017, bluck_what_2019, chen_galaxy_2024, figueira_quiescent_2024}.

Stronger observational constraints on the statistics of dwarf galaxy sizes and morphologies would help to improve models of galaxy formation. In state-of-the-art models, the `macroscopic' parameters associated with the galactic baryon cycle\footnote{I.e.\ those that determine the efficiency of star formation and various types of feedback, averaged over $\gtrsim100$~pc scales.} are calibrated to the statistics of bright galaxies, such as galaxy luminosity or mass functions \citep{schaye_eagle_2015}. Such calibrations can be degenerate, even where multiple properties are used for calibration -- more than one set of model ingredients may reproduce the calibration data. In principle, degenerate models can be distinguished by their predictions for statistics that are not determined (at least directly) by the calibration, but which nevertheless depend strongly on the evolution of the baryon cycle. 
The relationship between size, mass, and morphology is one example -- these properties are sensitive to exactly where, when, and how the stars in a galaxy are created. That current models struggle to reproduce the observed mass dependence of morphology and size likely reflects uncertainties in their treatment of the baryon cycle \citep[e.g.][]{bottrell_galaxies_2017, furlong_size_2017, bluck_what_2019, bluck_are_2020, somerville_relationship_2025}. The baryon cycle in massive galaxies is relatively complex (involving star formation occurring under different conditions over a long period of time, influenced also by AGN feedback and mergers). In contrast, dwarf galaxies (especially in the field) may provide a cleaner, more direct probe of the fundamental processes involved in star formation and stellar feedback.

In this work, we use data from the DESI Legacy Imaging Survey (DR9 and DR10) to explore the dwarf galaxy size-mass relation and its dependence on the environment.
The paper is organized as follows. In Sect.~\ref{sec:data}, we describe the data selection and the catalog we used to construct our samples. Our method for obtaining size-luminosity and size-mass relations is presented in Sect.~\ref{sec:result}. We discuss environmental effects in Sect.~\ref{sec:envs} and \ref{sec:discussion}.
We summarize our work in Sect.~\ref{sec:summary}. 
We assume a flat $\Lambda$CDM cosmology, $\Omega_\mathrm{m}+\Omega_\Lambda=1$ with $\Omega_\mathrm{m}=0.3$ and a Hubble constant $H_0=100h \, \mathrm{km \, s^{-1}\, Mpc^{-1}}$ with $h=0.7$ \citep{spergel_first-year_2003}.

\section{Data and sample selection} \label{sec:data}

The DESI Legacy Imaging Survey, LS, provides deep photometry with a pixel size of $\sim 0.27$ arcsec in three bands \citep[$g$, $r$, and $z$;][]{dey_overview_2019}. Compared to the previous generation of all-sky surveys (SDSS and PanSTARRS), DESI-LS is better suited to low-mass and low-surface brightness galaxy studies, with a surface brightness limit of $\sim$29 mag/arcsec$^2$ \citep[][]{tanoglidis_shadows_2021, roman_discovery_2021, martinez-delgado_hidden_2023, miro-carretero_extragalactic_2024}. 

\subsection{Sample selection \label{sec:qualitycut}}

We use LS Data Release 9 (DR9) and Data Release 10 (DR10) to construct samples of galaxies suitable for measuring the size-mass relation. DR10 expands on DR9 by incorporating additional southern sky coverage, including imaging from the Dark Energy Survey (DES), yielding a larger number of galaxies with photometric redshifts. However, the set of galaxies with spectroscopic redshifts remains the same as in DR9. Therefore, we construct two separate samples. The first sample, which we refer to as \sz\/, is based on spectroscopic redshifts using DR9 photometry matched with SDSS DR16 spectra. This sample contains observations of both the northern and southern sky. 
For galaxies without spectroscopic observations, we use DR10 photometry to construct another sample based on recent, high-quality photometric redshift measurements \citep[computed using $g$, $r$, $i$, $z$, $W1$, and $W2$ bands;][]{wen_catalog_2024}, which we call \pz\/.

The starting point for both samples is a set of photometric quality cuts to remove the galaxies with bad photometry. These cuts are the same as those used in the target selection for the DESI Bright Galaxy Survey, as outlined by \citet{ruiz-macias_characterizing_2021}.

For galaxies in the \sz{} sample, we apply cuts on the quality of the SDSS redshifts as follows: {\fontfamily{cmtt}\selectfont ZWARNING = 0 or 16,} {\fontfamily{cmtt}\selectfont PLATEQUALITY = `good',} {\fontfamily{cmtt}\selectfont SPECPRIMARY > 0,} {\fontfamily{cmtt}\selectfont CLASS = `GALAXY'.} Requiring
{\fontfamily{cmtt}\selectfont ZWARNING == 0} selects spectra with no known problem\footnote{The definitions of the SDSS spectral quality flags can be found at \url{https://www.sdss4.org/dr16/algorithms/bitmasks/#ZWARNING}}.  
The requirement {\fontfamily{cmtt}\selectfont PLATEQUALITY == `good'} ensures that the spectra meet the standard for high scientific quality\footnote{We have followed the recommendations for selecting good SDSS spectra, which can be found at \url{https://www.sdss4.org/dr16/spectro/catalogs/#Selectinggoodspectra}}. {\fontfamily{cmtt}\selectfont SPECPRIMARY > 0} removes duplicate spectra. 
We match \sz{} galaxies to the GALEX-SDSS-WISE Legacy Catalog \citep[GSWLC-X2;][]{salim_galexsdsswise_2016, salim_dust_2018}, which provides stellar masses ($M_\star$) based on SED fitting. The error on $M_\star$ ranges from 0.03 dex for passive galaxies to 0.10 dex for the most active star-forming galaxies.
We limit our sample to the redshift range of the GSWLC catalog, $0.01 < z < 0.3$.
In addition, we apply an angular size cut of $R_\mathrm{eff} > 1.5$ arcsec, to remove galaxies with sizes comparable to the SDSS point spread function (PSF).
We also apply an $r$-band magnitude cut ($r < 17.77$) and a surface brightness cut ($\mu_r < 23$), following \citet{shen_size_2003}.
In total, our \sz{} sample contains 540,228 galaxies.

Since SDSS spectroscopy is limited to  $r<17.77$, many faint galaxies observed in the LS do not have an SDSS spectroscopic redshift \citep{Strauss2002AJ....124.1810S}. Therefore, we construct the \pz{} sample from the LS DR10 matched with the catalog provided by \citet[][WH2024]{wen_catalog_2024}, which contains photometric redshift and stellar mass estimates.
We apply the same quality cuts as \citet{ruiz-macias_characterizing_2021}, $R_\mathrm{eff} > 1.5$ arcsec, and $0.01 < z < 0.3$ limits, as for the \sz\/ sample.
We exclude galaxies with $r > 23.9$, the $r$-band magnitude limit reported in \citet{dey_overview_2019}, and with $\mu_r > 29$ as reported in \citet{martinez-delgado_hidden_2023}.
After these selections, there are 9,513,732 galaxies in our \pz{} sample.

\subsection{Size, absolute magnitude, and mass measurement \label{sec:params}}

The primary measure of size we use in this study is the effective radius ($R_\mathrm{eff}$, half-light radius) provided in the LS catalog (as \texttt{SHAPE\_R}).
The $R_\mathrm{eff}$ provided by LS is obtained using Tractor, which fits a single morphological model across all bands ($g$, $r$, and $z$ for DR9; $g$, $r$, $i$, and $z$ for DR10).
However, $R_\mathrm{eff}$ in the LS refers to the circularized average radius. 
Therefore, we compute an equivalent semi-major axis effective size ($R_\mathrm{maj})$ using $R_\mathrm{eff}$ and the shape parameters (\texttt{SHAPE\_E1} and \texttt{SHAPE\_E2}) also provided by LS\footnote{The equations for computing the major axis effective radius from the circular effective radius are given at \url{https://www.legacysurvey.org/dr10/catalogs/\#toc-entry-4}.}.
In Fig.~\ref{fig:data_dis}, panels (a) and (b) show the distributions of $R_\mathrm{eff}$ and $R_\mathrm{maj}$ for our samples. As expected, the distribution of $R_\mathrm{maj}$ is broader than that of $R_\mathrm{eff}$.
Panel (c) shows the distribution of surface brightness within $R_\mathrm{eff}$, which indicates that the \pz{} sample contains more low-surface-brightness galaxies.

To obtain absolute magnitudes, we corrected for Galactic dust extinction using {\texttt{MW\_TRANSMISSION\_x}} provided in the LS catalog, applied as $m_x = 22.5-2.5\log_{10}(F_x/\tau_x)$, where $m$ is the apparent magnitude, $F$ is flux in units of nanomaggies, $\tau$ is the Milky Way transmission, and $x$ denotes the $g, r,$ and $z$ bands. From this corrected apparent magnitude, we use the spectroscopic redshift or photometric redshift to compute distance and absolute magnitude.
We also apply a $K$-correction using the python package {\texttt{kcorrect}} \citep{blanton_k_2007}. We use the {\texttt{desi\_x}} filters for $K$-correction.
In Fig.~\ref{fig:data_dis}, panels (d) and (e) show the distributions of $r$-band apparent magnitude and absolute magnitude for our samples. The sharp edge of the \sz{} sample around $r\approx17.77$ is due to the faint limit of SDSS spectroscopy \citep{abazajian_seventh_2009}. The advantage of the \pz{} sample is that it includes many more faint galaxies, with an approximate faint limit $M_r > -15$.

For the \pz{} sample, we use the stellar mass provided by WH2024, with an estimated uncertainty of approximately 0.15 dex \citep{yuan_more_2023, wen_catalog_2024}. 
The distribution of $M_\star$ is shown in Fig.~\ref{fig:data_dis}(f). By using the \pz{} sample, we can extend our investigation of sizes to galaxies with $M_\star \approx 10^{8}\, \msun{}$.

\begin{figure}
    \centering
    \includegraphics[width=\columnwidth]{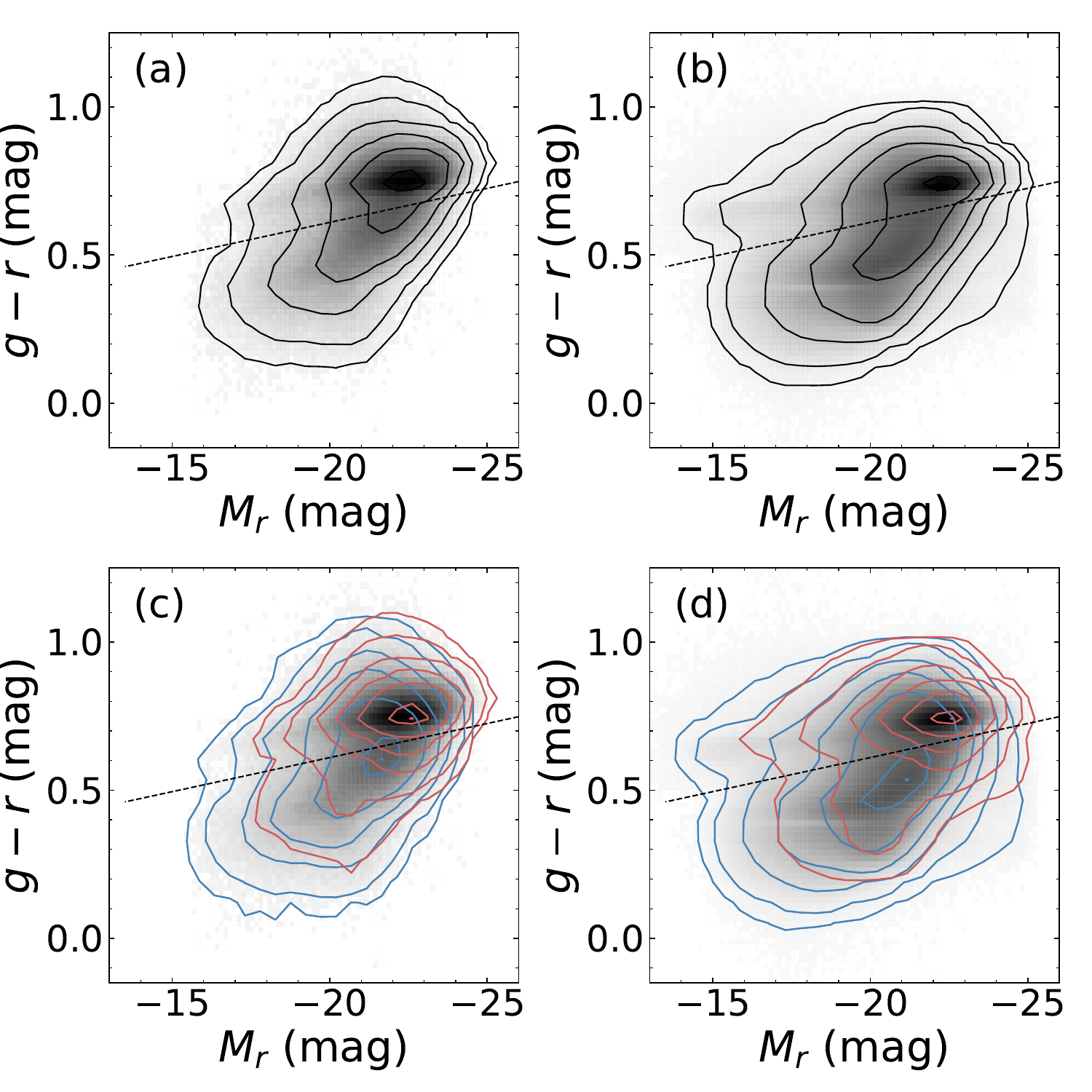}
    \caption{Upper row: color magnitude diagrams of the \sz\/ sample (panel a) and the \pz\/ sample (panel b). 
    Lower row:  contours show the distribution of galaxies with $n>2.5$ (red) and $n<2.5$ (blue) for the \sz\/ sample (panel c) and the \pz\/ sample (panel d).
    All the contours are drawn at [0.1, 0.5, 3, 10, 30, 80, 99]\% of the peak density.
    The dashed line indicates a fiducial separation between red sequence and blue cloud galaxies, $g-r = -0.023\times M_r+0.15$, determined by the eye. Although most galaxies with high S\'ersic index are located in the red sequence, the two definitions of the type clearly select different samples of galaxies.}
    \label{fig:cmd}
\end{figure}

\begin{figure}
    \centering
    \includegraphics[width=\columnwidth]{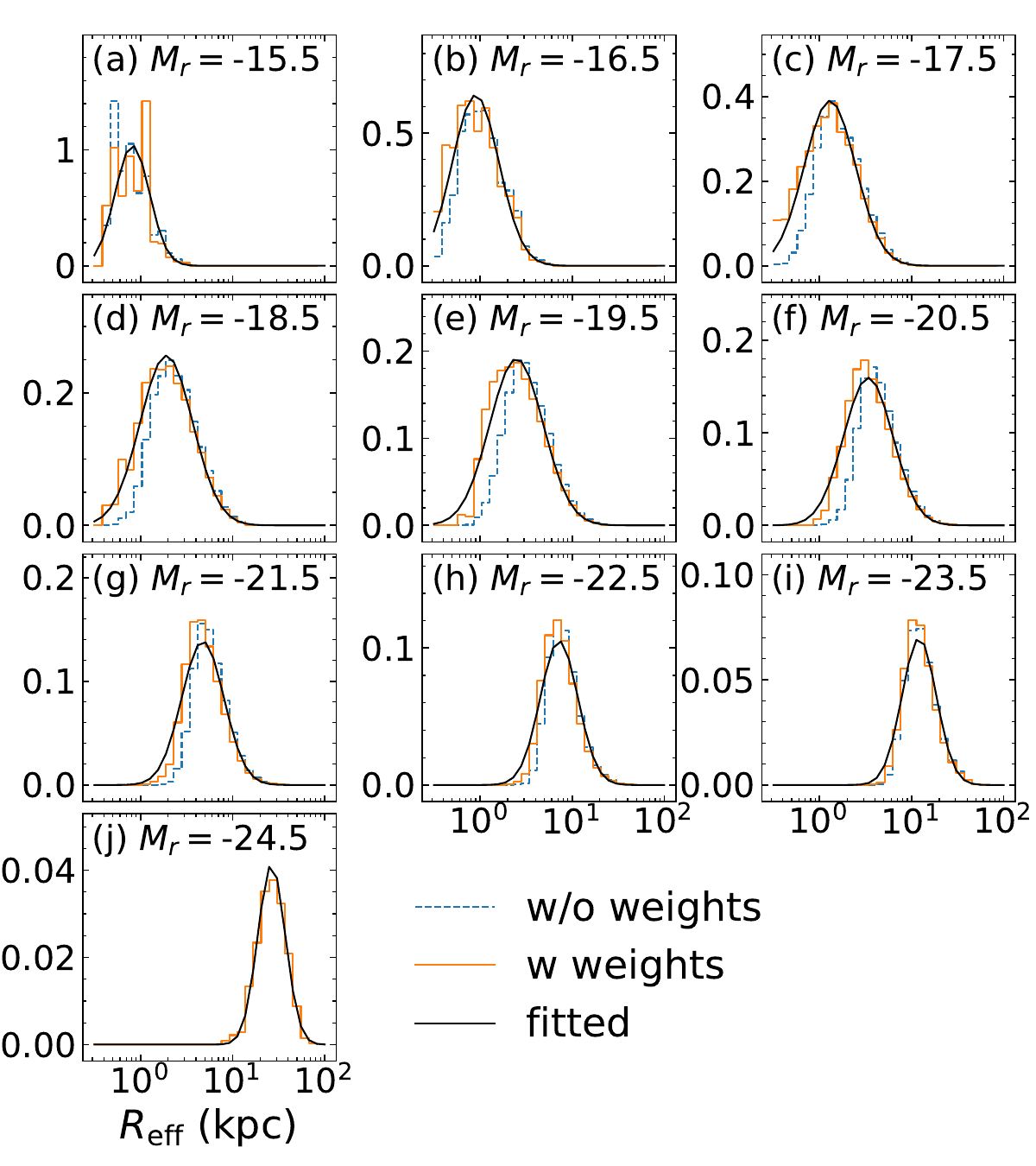}
    \caption{Distribution of effective semi-major axis for \sz\/ galaxies. The blue-dashed lines show the distribution without applying $V/V_\mathrm{max}$ weighting. The solid orange lines show the weighted distribution. The black lines show the results of maximum likelihood fits to the weighted distributions.}
    \label{fig:likelihood}
\end{figure}

\subsection{Volume correction \label{sec:Vcorrect}}

We follow \citet{shen_size_2003} by calculating a volume correction weight for each galaxy in our sample and reporting sample statistics based on those weights. This contrasts with the approach of restricting the volume of the sample to ensure completeness at a given magnitude \citep[e.g.][]{van_der_wel_3d-hstcandels_2014,mowla_mass-dependent_2019, chen_galaxy_2024}.

We first consider the effect of the flux limit \citep{schmidt_space_1968}. For each galaxy, the corresponding maximum luminosity distance ($d_{L,\mathrm{max}}$) for selection in to our sample is 
\begin{equation}
    \label{eq:dL}
    d_{L,\mathrm{max}} = 10^{-0.2(M_r-m_\mathrm{lim}-5)},
\end{equation}
where $M_r$ is the $r$-band absolute magnitude and $m_\mathrm{lim}$ is the limiting apparent magnitude. The value of $m_\mathrm{lim}$ for the \sz{} and \pz{} samples is 17.77 and 23.9, respectively.
From $d_{L,\mathrm{max}}$, we can compute the corresponding redshift, $z_\mathrm{d}$.

The surface brightness limit of the survey also imposes a redshift limit,
\begin{equation}
    \label{eq:zmu}
    z_\mathrm{max,\mu} = (1+z)10^{(\mu_\mathrm{lim}-\mu_r)/10}-1,
\end{equation}
where $\mu_\mathrm{lim}$ is the surface brightness limit and $\mu_r$ is the average surface brightness within $R_\mathrm{eff}$. We use $\mu_\mathrm{lim}= 23$~mag/arcsec$^2$ for the \sz{} sample \citep{shen_size_2003}  and $\mu_\mathrm{lim}=29$~mag/arcsec$^2$  for the \pz\/ sample, following \citet{martinez-delgado_hidden_2023}.

In addition, we exclude galaxies with $R_\mathrm{eff}<1.5$ arcsec, the limit imposed by the finite size of the LS PSF. 
To account for this size cut, we compute the minimum size limit following Equation 6 in \citet{shen_size_2003}:
\begin{equation}
    \frac{d_{A,\mathrm{max}}}{d_{A(z)}} = \frac{R_\mathrm{eff}}{R_\mathrm{PSF}},
\end{equation}
where $d_{A,\mathrm{max}}$ is the angular diameter distance at the sample maximum redshift, $d_{A(z)}$ is the angular diameter distance of the galaxy, and $R_\mathrm{PSF}$ is the PSF size limit, i.e.\ the smallest galaxies permitted in the sample. Having determined $d_{A,\mathrm{max}}$, we can compute the corresponding redshift, $z_A$.

The final redshift range associated with a given galaxy is therefore [0.01, min($z_d$, $z_\mu$, $z_A$, 0.3)]. The corresponding volume is computed using the comoving distance ($d_c$) of the minimum and maximum redshift:
\begin{equation}
    V_\mathrm{max} = \frac{4\pi}{3}\left(d_c(z_\mathrm{max})^3-d_c(z_\mathrm{min})^3\right).
\end{equation}
Fig.~\ref{fig:data_dis} shows distributions of galaxy observables computed with and without the $1/V_\mathrm{max}$ weights. After the volume correction, the peaks of these distributions move to fainter magnitudes, reflecting the contribution of intrinsically faint galaxies missing from the sample at larger distances.
We neglect the $k$-correction when computing $V_\mathrm{max}$; our reasons for doing so are given in Appendix~\ref{appendix:kcorrection}.

\section{Results}
\label{sec:result}

\begin{figure*}
    \centering
    \includegraphics[width=\textwidth]{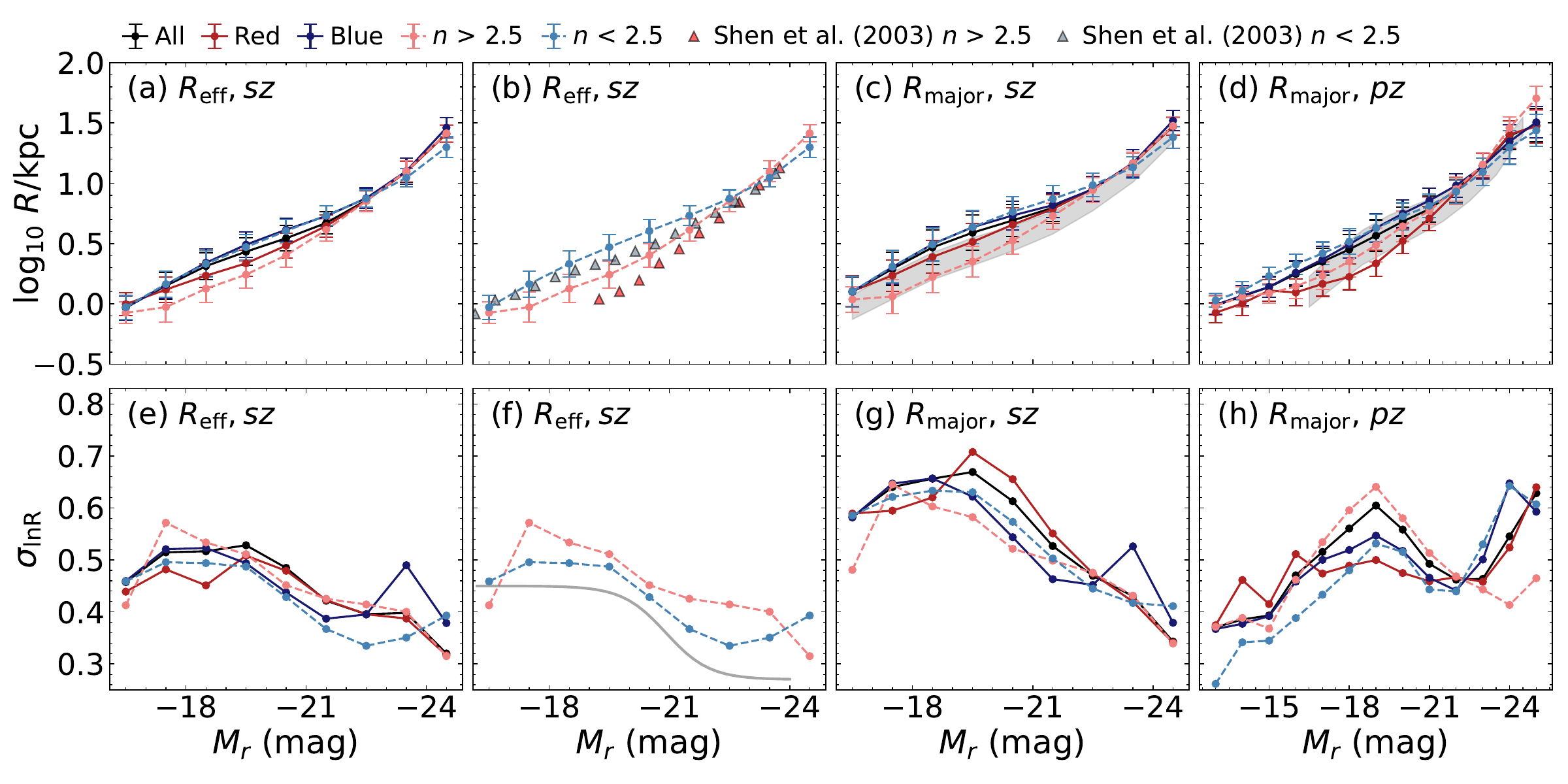}
    \caption{Magnitude--size relations for our samples. From left to right, panels show: (a) the $M_r$ vs. circular effective radius $R_\mathrm{eff}$ for our \sz{} sample; (b) comparison of the results in panel (a) to those of \citet{shen_size_2003}; (c) the  $M_r$ vs. semi-major axis radius $R_\mathrm{maj}$ for our \sz{} sample; (d) $M_r$ vs. $R_\mathrm{maj}$ for our \pz{} sample. The black circles show results for all the galaxies in a particular sample. Dark red and blue circles show results for the color-selected red and blue subsamples, respectively. Pink and light blue circles show results for the high and low S\'ersic index subsamples, respectively. 
    The error bars represent the dispersion of the maximum likelihood fits.
    In panel (b), the red and gray triangles are data points from \citet{shen_size_2003}.
    For ease of comparison, the shaded gray region in (c) reproduces the $R_\mathrm{eff}$ relation for the \sz{} sample from panel (a). 
    The shaded gray region in panel (d) reproduces the result for all the galaxies in the \sz{} sample using $R_\mathrm{maj}$ from panel (c).
    We ensure that each $M_r$ bin contains at least 100 galaxies.
    The bottom panels (e-h) show how the corresponding dispersion around each relation varies with $M_r$. In panel (f), the dispersion found by \citet{shen_size_2003} is shown with a thin gray line. 
    }
    \label{fig:Mr_size}
\end{figure*}

In this section, we explain how we determine the average size of galaxies in a given magnitude or stellar mass bin, from which we derive the size-magnitude and size-mass relations of our \sz{} and \pz{} samples. In addition, we separate our galaxies into `morphological' early- and late-type subsets using the S\'ersic index provided in the LS catalog \citep[obtained from fitting to the 2d surface brightness distribution, using \texttt{Tractor};][]{lang_tractor_2016}. We define early-type galaxies as having $n > 2.5$ and late-type galaxies as having $n < 2.5$, following \citet{shen_size_2003}. For simplicity, in the following text, we refer to galaxies with $n > 2.5$ as elliptical galaxies and those with $n < 2.5$ as disk galaxies.
However, the disk structure of faint galaxies may be less prominent due to observational limitations, and many of these galaxies may not have disks at all. A definition of late and early types based on the Sérsic index is thus less straightforward to interpret in terms of visual morphology at fainter absolute magnitudes. We also classify galaxies based on $g-r$ color and $M_r$. This separation on the color-magnitude diagram (CMD) provides a different way to distinguish between late and early types. At brighter magnitudes, morphologically elliptical galaxies predominantly occupy the red sequence, while disk galaxies are typically (but not exclusively) found in the blue cloud \citep{weinmann_properties_2006}.
Fig.~\ref{fig:cmd} shows the CMDs of the \sz{} and \pz{} samples, illustrating the significant differences between the morphological and color-based subsets of our sample.
At the faint end, we note that there are more red galaxies in \pz{} sample. This may reflect a photo-z bias, whereby faint blue galaxies are systematically assigned larger redshifts and therefore appear brighter \citep[e.g.,][]{Ilbert2006A&A...457..841I, Dahlen2013ApJ...775...93D, 2022ARA&A..60..363N}. Consequently, some low-mass galaxies may be underrepresented in our photo-z sample.

\subsection{Fitted size using maximum likelihood}
\label{sec:size_like}

Fig.~\ref{fig:likelihood} illustrates the distribution of sizes and typical sizes for galaxies in different ranges of magnitude. We obtain these distributions by weighted maximum likelihood fitting for galaxies in bins of either $M_r$ or $M_\star$.
Following \citet{shen_size_2003}, we assume a log-normal distribution:
\begin{equation}
    L=\prod_{i=0}^n \left\{ \frac{1}{x_i \sigma \sqrt{2\pi}} \exp \left[ -\frac{(\ln x_i-\mu)^2}{2\sigma^2} \right] \right\}^{w_i},
\end{equation}
where $x_i$ is size of each galaxy, $\mu$ is the median of the size, $\sigma$ is the dispersion of the size, and $w_i$ is the $1/V_\mathrm{max}$ weighting of each galaxy.
The weighted log-likelihood is as follows:
\begin{equation}
    \log L = \Sigma_{i=1}^n \, w_i \left[ -\log(x_i \sigma \sqrt{2\pi})-\frac{(\ln x_i-\mu)^2}{2\sigma^2}\right].
\end{equation}
We then use the $\mu$ and $\sigma$ that maximize the likelihood as the representative size and dispersion for the corresponding bin.

\subsection{$M_r$ and size} \label{sec:Mrsize}

Fig.~\ref{fig:Mr_size} begins our exploration of the relationship between absolute magnitude (in the $r$-band, $M_r$) and physical size. Panels (a) and (c) show the $M_r$ -- size relations using $R_\mathrm{eff}$ and $R_\mathrm{maj}$, respectively.
Elliptical and disk galaxies are plotted with light red and blue circles, respectively.
Galaxies classified as red or blue based on their location on the CMD are shown with dark red and blue circles, respectively.
In general, brighter galaxies tend to have larger sizes, as reported in previous literature \citep[e.g.,][]{shen_size_2003, van_der_wel_3d-hstcandels_2014, mowla_mass-dependent_2019}. The relation steepens for galaxies brighter than $M_r < -22$. The various subsets of our sample all produce very similar relations in this regime.
The differences between subsets are more prominent at $M_r > -22$, where the sizes of galaxies in the blue/disk subsets are larger than those of galaxies in the red/elliptical subsets, at fixed $M_r$.

Fig.~\ref{fig:Mr_size} panel (c) shows that, at a given magnitude, average sizes defined by $R_\mathrm{maj}$ are larger than those defined by $R_\mathrm{eff}$ \citep[e.g.][]{dutton_evolution_2011}. We find the difference between these two measures of size is greater for fainter galaxies.
This is expected because $R_\mathrm{maj}$ differs most from $R_\mathrm{eff}$ for highly inclined disk galaxies and less for elliptical galaxies; the ratio of elliptical galaxies to disk galaxies in all the samples increases at brighter $M_r$.

In Fig.~\ref{fig:Mr_size} panel (b), we overplot size-mass relations from \citet{shen_size_2003} for galaxies with $n > 2.5$ and $n < 2.5$. 
In general, our relations are in good agreement with those of \citet{shen_size_2003}, especially for disk galaxies. Since \citet{shen_size_2003} uses the Petrosian half-light radius as their measure of size, Fig.~\ref{fig:Mr_size} panel (b) is close to a like-for-like comparison.
For bright galaxies ($M_r \leq -22$), we find sizes comparable to those reported by \citet{shen_size_2003}. 
For faint galaxies, we find sizes that are $\sim0.2$ dex larger than those reported by \citet{shen_size_2003}.
This could be due to the different methods used to measure the half-light radius (see appendix \ref{appendix:ls_sdss}, Fig.~\ref{fig:ls_sdss} panel b).
\texttt{Tractor} determines sizes by fitting an ellipsoidal model to the galaxy image, while \citet{shen_size_2003} uses the Petrosian half-light radius measured in circular apertures. 
However, the differences between these two definitions of radius are not expected to be as large as those we find \citep[e.g. a difference of $\sim0.15$ dex is reported by][]{dutton_evolution_2011}. Another contributing factor could be that LS detects more large, faint galaxies with lower surface brightness \citep[e.g.][]{walmsley_deep_2023}. 
Alternatively, we cannot exclude the possibility that this effect is due to systematic errors in the LS size measurements for fainter galaxies \citep{moustakas_siena_2023, darragh-ford_target_2023}; we discuss this further in the appendix (Fig.~\ref{fig:ls_sdss} (d)).

Fig.~\ref{fig:Mr_size} panel (d) shows the size-magnitude relation for the \pz\/ sample. 
The deeper \pz\/ data allow us to extend the relation two magnitudes fainter than that of the \sz\/ sample. 
We overlay the result from the \sz{} sample (gray shaded gray region). The sizes obtained from the \pz{} sample are approximately 0.1 dex larger but remain within the dispersion of the distribution.
We note that our angular size cut to exclude unresolved objects could bias the results for the faintest bins. A detailed discussion of the $M_r$--size relation of dwarf galaxies is given in Sect.~\ref{sec:Mr_size_dw}.

For comparison with the literature, we fit a straight line to our size-magnitude relations,
\begin{equation} 
\log_{10} R/\mathrm{kpc} = a\times M_r+b,
\label{eq:sizemag}
\end{equation}
where $a$ represents the slope of the relation. The best-fit values are reported in Table~\ref{tab:size_Mr}. The results show that the relation for red/elliptical galaxies is steeper than that for blue/disk galaxies, in agreement with previous studies \citep{shen_size_2003}.

The lower panels of Fig.~\ref{fig:Mr_size} show the dispersion of size ($\sigma_{\ln R}$) as a function of $M_r$. For galaxies with $M_r < -16$, the dispersion decreases with increasing brightness.
This is also consistent with previous studies \citep{shen_size_2003,
bernardi_systematic_2014, cebrian_effect_2014}. 
In Fig.\ref{fig:Mr_size}, panel (f), we overlay the $\sigma_{\ln R}$ from \citet{shen_size_2003} as a gray line. The dispersion in our work is larger by approximately 0.1~dex. This could be due to the larger number of galaxies in LS
or to differences between the size measurements in LS and SDSS (see Appendix~\ref{appendix:ls_sdss}).
The dispersion in $R_\mathrm{maj}$ is larger than that in $R_\mathrm{eff}$, but has a similar trend.
Toward the bright end, where most galaxies are elliptical, the difference between the two definitions is small. The dispersion increases at intermediate magnitudes according to both definitions. We believe this can be explained by the greater variety of galaxy structure at intermediate magnitudes (both overall morphology and the bulge fractions of disks). The increase is greater for $R_\mathrm{maj}$, which suggests an intrinsically larger dispersion in disks (perhaps reflecting the spread of angular momenta and halo concentration). The dispersion narrows again towards the faint limit of our spectroscopic sample ($M_r > -16$); central bulges are less prominent (or completely absent) in this regime, although it may also be that the largest galaxies fall below the surface brightness limit of our sample at the faintest magnitudes. The difference between the dispersions of $R_\mathrm{eff}$ and $R_\mathrm{maj}$ remains constant to the faint limit of our spectroscopic sample. 
We consider the correspondence between galaxy structure and size dispersion in more detail in Appendix~\ref{appendix:sigma_n}.

\subsection{$M_\star$ vs. size}

\begin{figure}
    \centering\includegraphics[width=1\columnwidth, trim=30 10 0 0]{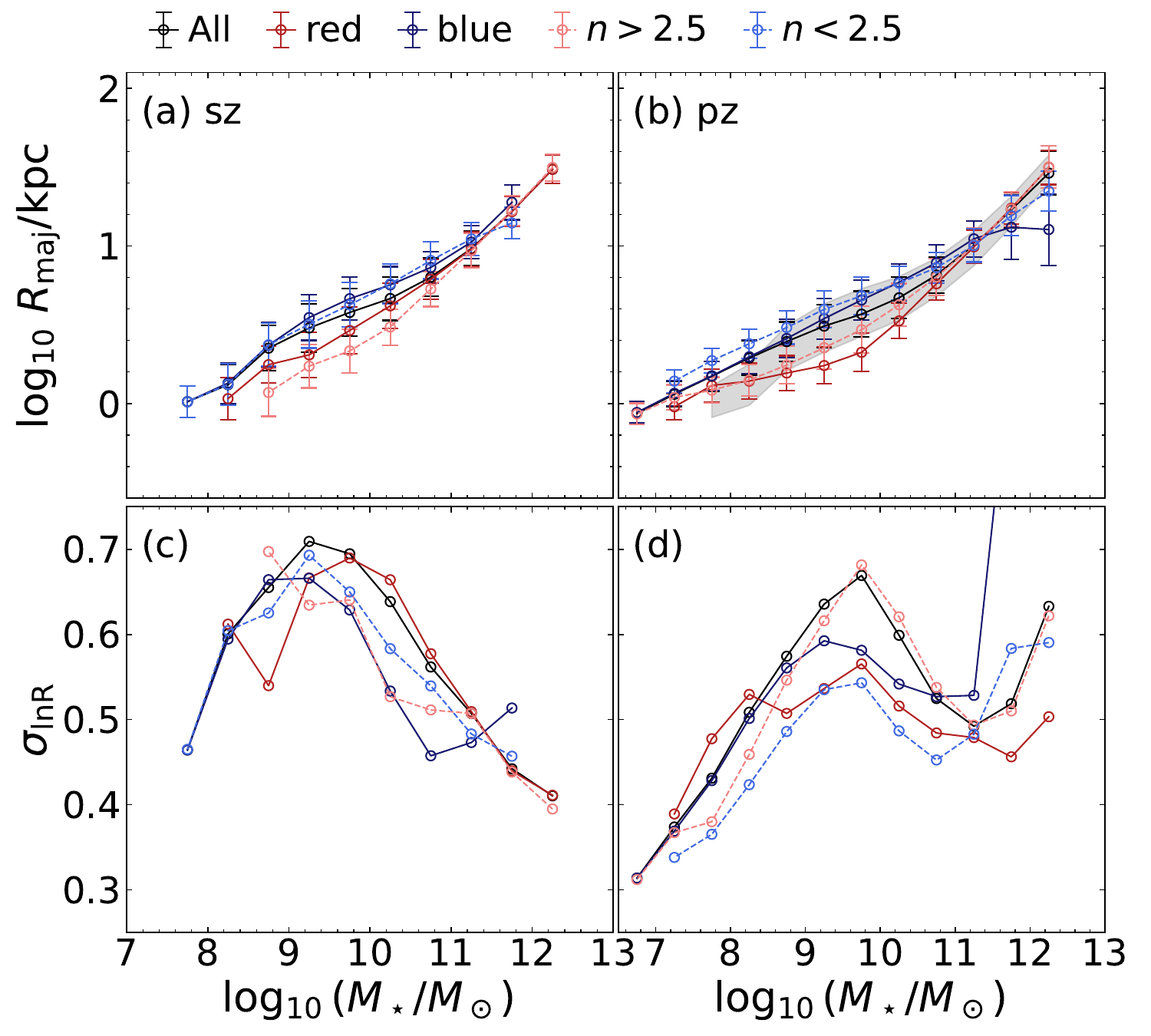}
    \caption{Size-mass relation for the (a) \sz\/ and (b) \pz\/ samples. Masses for \sz\/ galaxies are taken from the GSWLC catalog \citep{salim_galexsdsswise_2016}. Masses for \pz\/ galaxies are taken from WH2024.
    The line styles and symbols have the same meaning as those in Fig.~\ref{fig:Mr_size}. The error bars represent the dispersion, $\sigma$, of the log-likelihood.
    The bottom panels show $\sigma$ vs. $M_\star$ for (c) \sz\/ and (d) \pz\/ samples.
    The gray region in panel (b) reproduces the result for all the galaxies in the \sz{} sample from panel (a).}
    \label{fig:size_mass}
\end{figure}

Fig.~\ref{fig:size_mass} shows the size-stellar mass relations of the \sz\/ and \pz\/ samples.
The trends for different subsets are similar to those seen in the size-luminosity relation.
Again, the difference between blue/disk galaxies and red/elliptical galaxies is significant at stellar masses $8.5 \lesssim \log_{10}M_\star/M_\odot \lesssim 10.5$. In contrast, for galaxies with a mass greater than $\log_{10} M_\star/M_\odot \sim10.5$, the slopes of the size-mass relations are generally similar. 
We fit the slope of the size-mass relation with the function
\begin{equation}
    \log_{10} R/\mathrm{kpc} = a\log_{10} M_\star/\mathrm{M_\odot} + b.
\label{eq:sizemasslog}
\end{equation}
The results are listed in Table~\ref{tab:size_Mr}.
Overall, the logarithmic slope of the size-mass relation is $a \approx0.3$, steepening to between $0.4$ and $0.6$ for redder galaxies with $\log_{10}M_\star/M_\odot > 10.5$.
These results agree with previous studies at low redshift \citep[e.g.][]{shen_size_2003, van_der_wel_3d-hstcandels_2014, lange_galaxy_2016, mowla_mass-dependent_2019, chen_galaxy_2024, chamba_impact_2024, 2025arXiv251013719V}.
Although the sizes in the \pz{} sample are slightly larger than those from \sz{} at fixed mass, especially for galaxies with $\log_{10} M_\star/ \msun \leq 9$, the results from both samples are in good agreement overall.

\begin{figure}
    \centering
    \includegraphics[width=\linewidth]{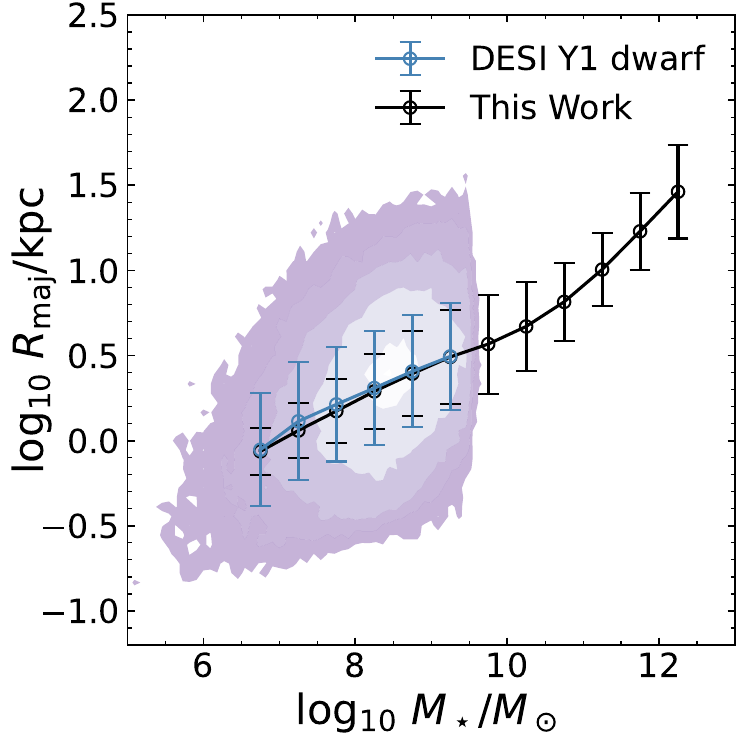}
    \caption{Size-mass relations for our \pz{} sample (black), based on photometric redshifts, and a sample of dwarf galaxies with spectroscopy from the DESI Y1 data release (blue). The error bars show the dispersion ($\sigma$) of the maximum likelihood fits to the size distribution at a given mass. The purple contours show the full distribution of the DESI Y1 galaxies.}
    \label{fig:desiy1}
\end{figure}

In Fig.~\ref{fig:desiy1} we compare the all-galaxy relation for our \pz{} sample, shown in \ref{fig:Mr_size}, to those of galaxies in the DESI Y1 data release. Specifically, we use the DESI Extragalactic Dwarf Galaxy Catalog (Manwadkar et al. 2025a,b in preparation\footnote{\url{https://data.desi.lbl.gov/doc/releases/dr1/vac/extragalactic-dwarfs/}}). DESI observes much fainter galaxies than the SDSS spectroscopic sample, so this comparison provides a test of the reliability of our results based on photometric redshifts. 
In order to compare with our volume-corrected results, we only use DESI Y1 galaxies with $z < 0.05$, to ensure the sample is complete at faint magnitudes.
We obtain $R_\mathrm{maj}$ for DESI Y1 galaxies from {\texttt{SHAPE\_R}} and {\texttt{BA}}, and the mass is from {\texttt{LOGM\_CIGALE}} \citep{boquien_cigale_2019, siudek_value-added_2024}.
Fig.~\ref{fig:desiy1} demonstrates good agreement between our photometric redshift sample and the smaller sample of comparably faint galaxies with DESI Y1 spectroscopy.

\subsection{Comparison with dwarf galaxies in the literature}
\label{sec:Mr_size_dw}

The deeper LS observations provide an opportunity to extend size-luminosity studies to fainter galaxies. \citet{brasseur_what_2011} combines the results from \citet{shen_size_2003} with size measurements for individual satellites of Andromeda and the Milky Way, to examine the form of the size-luminosity relation towards the faint end of the luminosity function. However, there is a significant gap in the data around $M_V\sim-15$, which we are now able to address with our analysis of the LS.

\begin{figure}
    \centering
    \includegraphics[width=\columnwidth, trim=30 10 0 10]{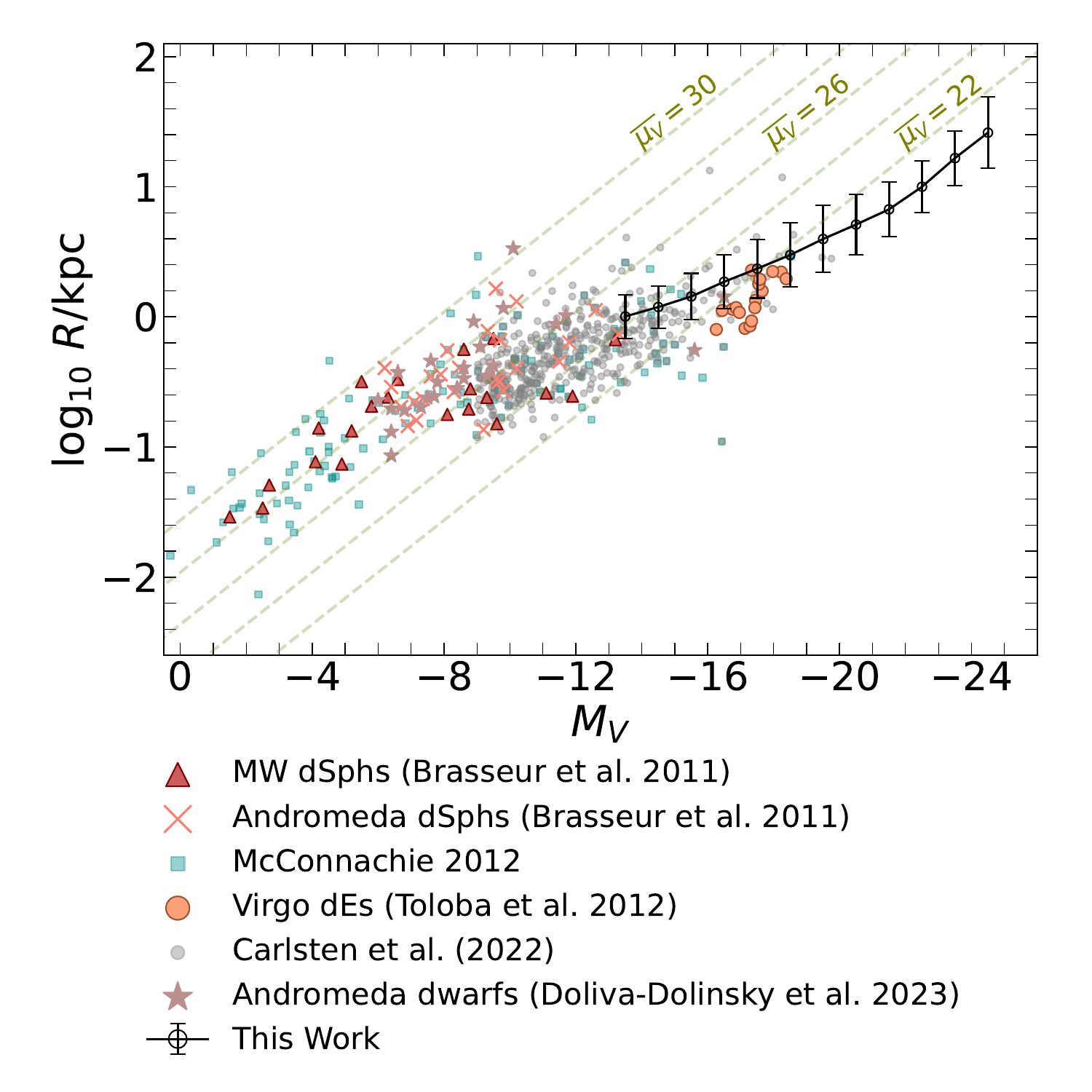}
    \caption{The relation between size and $V-$band magnitude for our \pz\/, sample (black circles, with error bars showing the 10th to 90th percentile range).  We compare with results from the dwarf galaxy literature: 
    Milky Way dSphs \citep[red triangles; ][]{brasseur_what_2011}; Andromeda dSphs \citep[orange crosses; ][]{brasseur_what_2011}; dwarf galaxies in the Local Group \citep[blue squares; ][]{mcconnachie_observed_2012}; Virgo dEs \citep[orange circles; ][]{toloba_formation_2012}; satellite dwarf galaxies reported by the ELVES project \citep[gray dots; ][]{carlsten_exploration_2022}; and Andromeda dwarf galaxies from \citet[][stars]{doliva-dolinsky_pandas_2023}. The dashed green lines show the surface brightness limit in units of mag arcsec$^{-2}$.
    }
    \label{fig:MVsize}
\end{figure}

Much of the literature on faint dwarf galaxies reports $V-$band magnitudes. We use LS $g$ and $r$ to estimate $V$ following Eq.~2 in \citet{carlsten_structures_2021}\footnote{This relation is based on the transformation for SDSS filters \citep{lupton_experience_2005}: \url{https://www.sdss3.org/dr8/algorithms/sdssUBVRITransform.php#Lupton2005}.}:
\begin{equation}
\label{eq:vband}
    V = g-0.5784\,(g-r)-0.0038.
\end{equation}
We note this equation is for the SDSS $g$ and $r$ filters; here we use the LS $g$ and $r$ magnitudes without first correcting for the small difference between the LS and SDSS photometric systems. In Appendix~\ref{appendix:ls_sdss}, we explain this choice in more detail and consider the systematic uncertainty associated with it -- this may be up to $\sim0.5$ mag for the brightest galaxies in our data, but is much smaller for dwarf galaxies ($M_V > -18$). This difference is negligible in comparison to the $\sim24$~mag range of Fig.~\ref{fig:MVsize}.

Fig.~\ref{fig:MVsize} shows the sizes we obtain from LS overlaid with data from nearby known dwarf galaxies, including the dSphs satellites of the Milky Way \citep{brasseur_what_2011} and  Andromeda \citep{brasseur_what_2011, doliva-dolinsky_pandas_2023}; dEs in the Virgo cluster \citep{toloba_formation_2012}; dwarf galaxies in the Local Group \citep{mcconnachie_observed_2012}; and dwarf satellites in the ELVES sample \citep{carlsten_exploration_2022}. The LS survey makes good progress in bridging the gap between large statistical samples and these smaller studies of nearby dwarf galaxies.
Visual comparison of our average relation with these literature data suggests that the slope of the size-luminosity relation becomes shallower in the range $-16 < M_V < -8 $ and then steepens again at fainter magnitudes, $M_V > -5$. 
However, we note that the apparent steepening of the slope at the faint end may be an artifact of the surface-brightness limit of the available data.
In addition, in this faint regime, the available samples are dominated by Local Group dwarf galaxies, including many satellites. The sizes of some of these galaxies may have been affected by tidal forces \citep{Asali2025arXiv250925335A}.

\begin{figure}
    \centering
    \includegraphics[width=\columnwidth, trim=10 10 0 10]{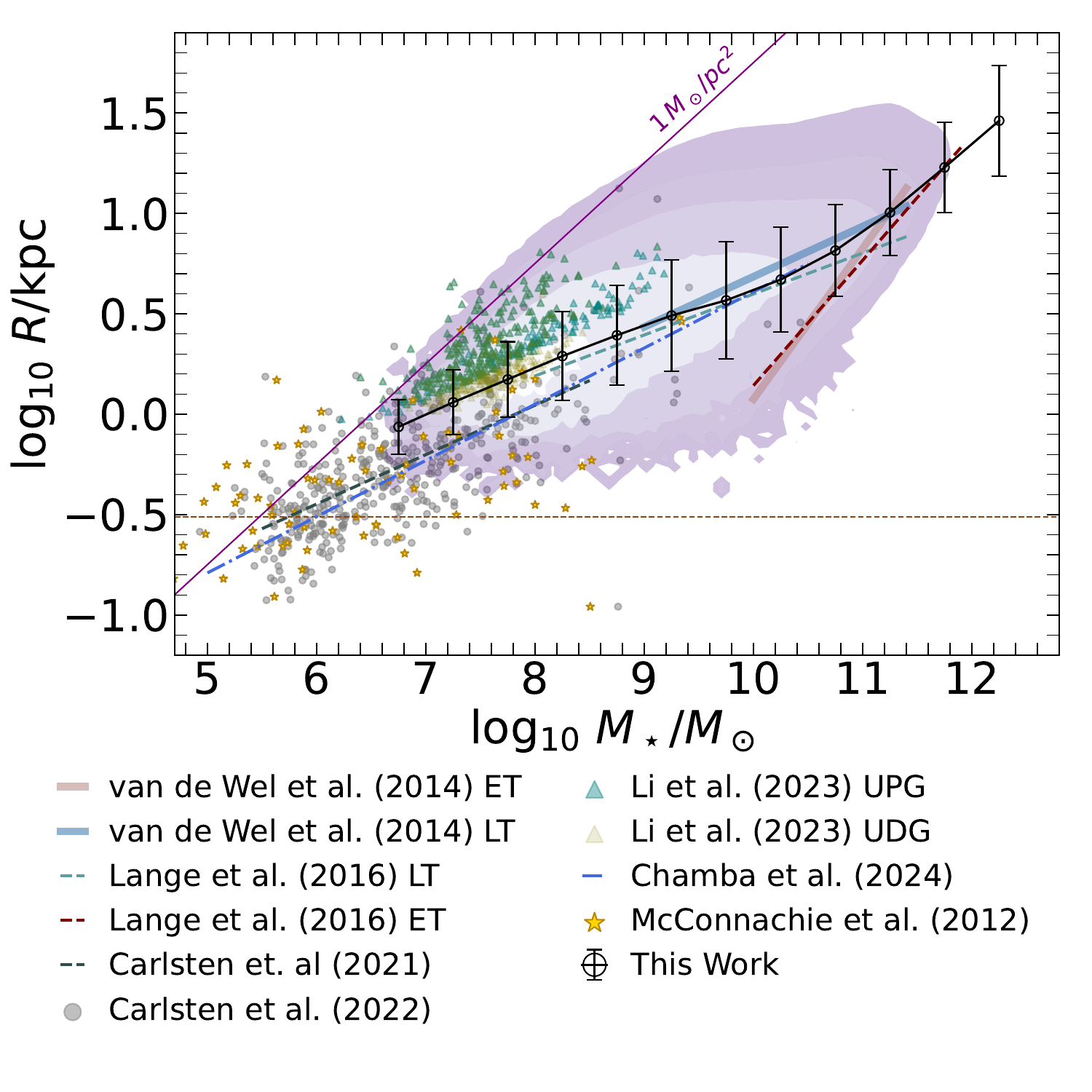}
    \caption{Size-mass relations from this work (black points with error bars showing the 10th to 90th percentile range) compared to others in the literature (line styles and colors shown in the legend):  
   \citet{van_der_wel_3d-hstcandels_2014}, 
   \citet[][GAMA]{lange_galaxy_2016}; and \citet{chamba_impact_2024}. 
   Points show individual galaxies from \citet{mcconnachie_observed_2012},
   \citet[][ELVES]{carlsten_exploration_2022}
   and 
   \citet{li_beyond_2023}, the latter for ``ultra-puffy'' and ``ultra-diffuse'' galaxies as defined in that work.
  The purple contours show the distribution of our galaxies. 
  The purple line shows a constant stellar mass surface density of  $1\,M_\odot/\mathrm{pc}^2$.
  The orange horizontal line shows the lower size limit of our sample.}
    \label{fig:sizemass_overlay}
\end{figure}

Fig.~\ref{fig:sizemass_overlay} shows our size-mass relation overlaid with other average relations from the literature. Our results at high mass are consistent with those of \citet{van_der_wel_3d-hstcandels_2014} and \citet{lange_galaxy_2016}.
The size-mass relation for red galaxies in \citet{van_der_wel_3d-hstcandels_2014} and \citet{lange_galaxy_2016} appears to deviate from that of our full sample. This difference is not due to a lack of small red galaxies in our sample, but rather because those galaxies do not dominate the average.
For lower-mass galaxies, we compare with results from \citet{mcconnachie_observed_2012}, \citet{carlsten_structures_2021}, and \citet{chamba_impact_2024}. 
The sizes reported in those studies are more compact than our results, with a difference of up to 0.2 dex.
This could be due to the different approaches used to obtain the size: for instance, we use $R_\mathrm{maj}$ in our size-mass relation, whereas \citet{chamba_impact_2024} measure disk sizes based on the point at which the slope of the surface brightness profile changes. It could also be a sample selection effect: our sample includes both field and satellite galaxies, while \citet{carlsten_structures_2021} focuses on nearby satellite dwarf galaxies.
In addition, the angular size cut in our sample may bias the average size toward larger values.

Alternatively, our results may reflect a different combination of dwarf galaxy types in our sample, which includes a wide range of environments. For example, we also show in Fig.~\ref{fig:sizemass_overlay} the ultra-diffuse dwarf galaxies and ``ultra-puffy'' galaxies \citep[those $1.5\sigma$ above the average size-mass relation of][]{carlsten_structures_2021} reported by \citet{li_beyond_2023}. These galaxies lie above our relation and (by definition) significantly above the mean of the ELVES data; the ELVES sample itself contains very few galaxies of similar size. This suggests that our relation may reflect a larger average size at low stellar mass for galaxies in the volume we sample, compared to the dwarf satellite galaxies in ELVES.

\section{Environmental effects on the size-mass relation}
\label{sec:envs}

\begin{figure}
    \centering
    \includegraphics[width=0.98\columnwidth]{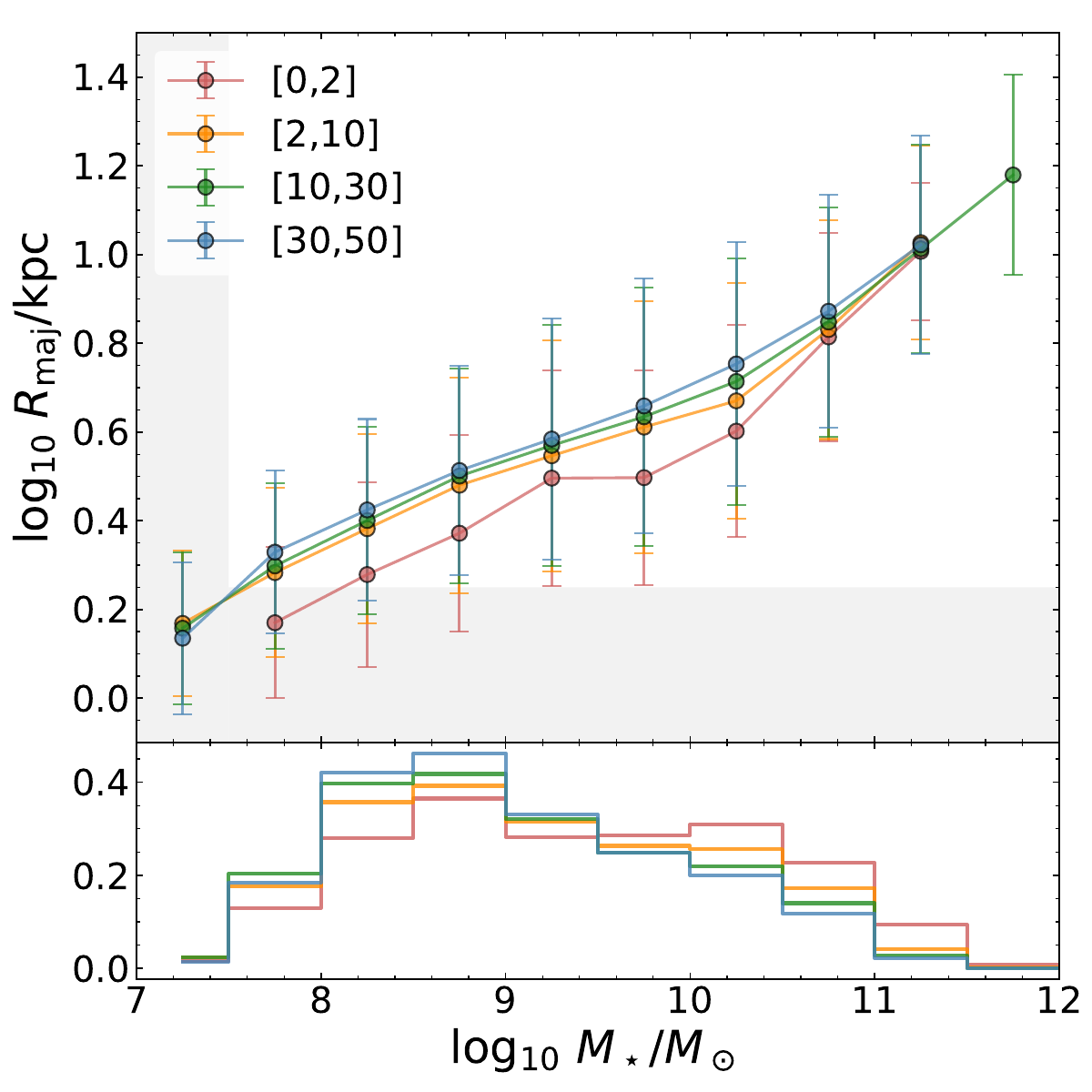}
    \caption{Size-mass relation for galaxies separated by $d_\mathrm{cl}$ (upper) and normalized distribution of galaxies (bottom). The galaxies with $d_\mathrm{cl}$ in [0, 2], [2, 10], [10, 30], and [30, 50] Mpc are plotted in red, orange, green, and blue, respectively. The error bars are the dispersion in the likelihood fitting. The gray area represents the incomplete region. We perform maximum likelihood fitting to determine the size only for mass bins that contain more than 50 galaxies. }
    \label{fig:dcl_all}
\end{figure}

\begin{figure*}
    \centering
    \includegraphics[width=\linewidth]{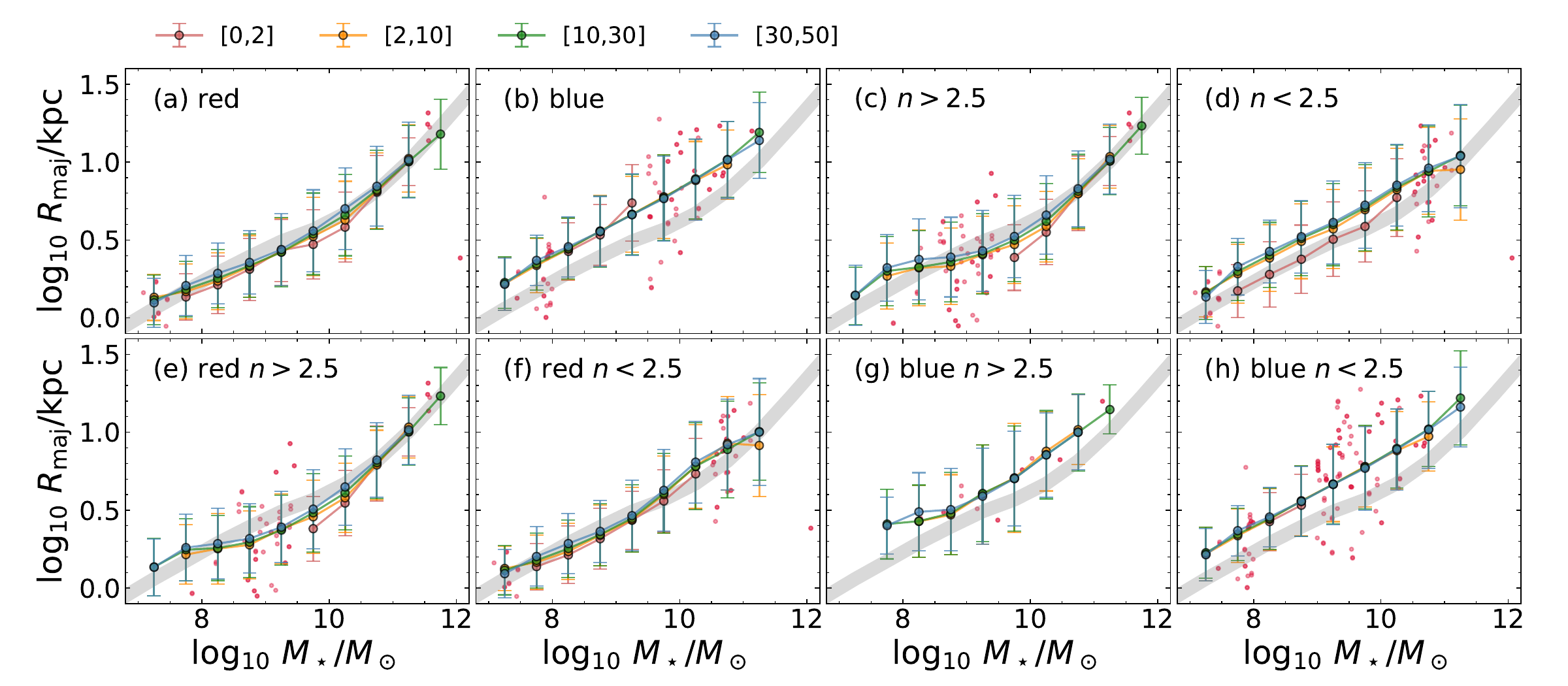}
    \caption{Size-mass relation for galaxies separated by $d_\mathrm{cl}$ for (a) red, (b) blue, (c) elliptical, (d) disk, (e) red elliptical, (f) red disk, (g) blue elliptical, and (h) blue disk galaxies. The galaxies with $d_\mathrm{cl}$ in [0, 2], [2, 10], [10, 30], and [30, 50] Mpc are plotted in red, orange, green, and blue, respectively. The error bars are the dispersion in the likelihood fitting. The gray thick lines show the size-mass relation for all the \pz{} samples. Only the mass bins containing more than 50 galaxies are plotted. The red points represent cluster galaxies with insufficient numbers to determine the mean size.}
    \label{fig:envs_all}
\end{figure*}

\begin{figure*}
    \centering
    \includegraphics[width=\linewidth]{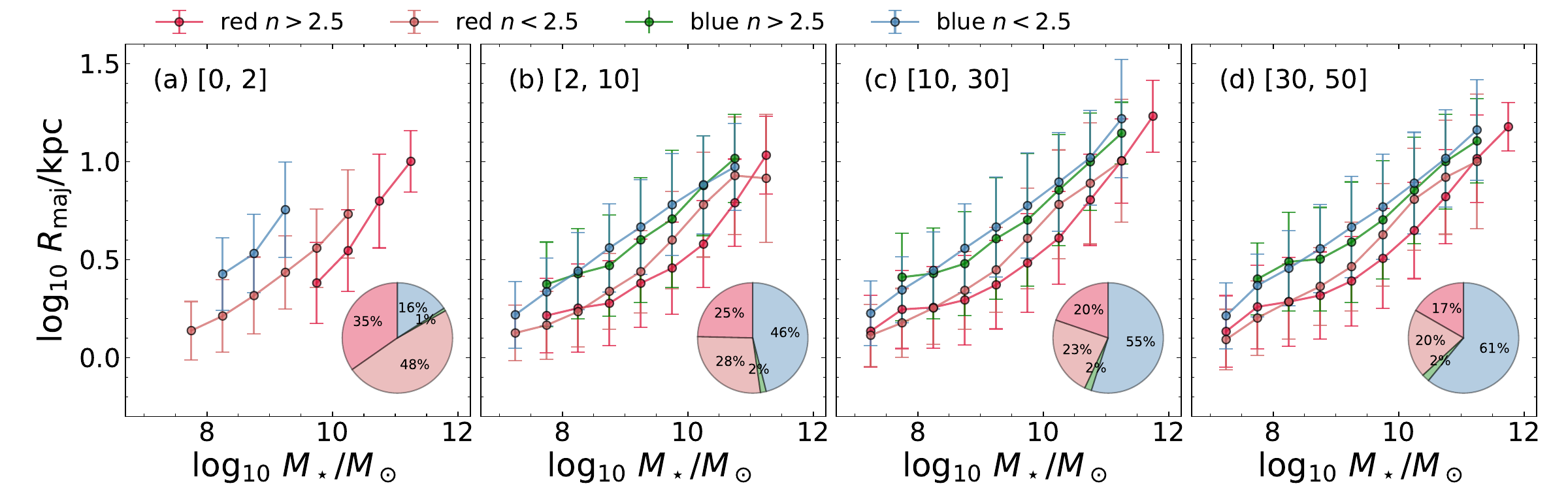}
    \caption{Size–mass relations for red $n>2.5$ (red), red $n<2.5$ (pink), blue $n>2.5$ (green), and blue $n<2.5$ (blue) galaxies in different environments (distance range in Mpc from the nearest galaxy cluster, shown in the top left of each panel). The pie charts show the fractions of each galaxy type in the corresponding environment. The error bars represent the dispersion of the likelihood fitting.}
    \label{fig:envs_morp}
\end{figure*}

Previous studies have come to a wide range of conclusions regarding the variation of typical galaxy sizes with cosmic environment. 
For example, \citet{huertas-company_no_2013}, \citet{cappellari_effect_2013}, and \citet{chen_galaxy_2024} find no or very subtle size differences for galaxies in different environments, while \citet{poggianti_superdense_2013} and \citet{cebrian_effect_2014} find significant differences.

To address this, we use the galaxy cluster catalog of WH2024 (based on the LS; see Sect.~\ref{sec:data}). The estimated lower mass limit of clusters in this catalog is $M_{500} \gtrsim 0.5 \times 10^{14}\,\msun$. For each galaxy in our dataset, we compute the 3-dimensional distance to the nearest cluster center ($d_\mathrm{cl}$) using photometric redshift.
Since our distances are computed using photometric redshifts, the precision of the environmental measure is limited by photo-$z$ uncertainties. Galaxies with large photo-$z$ errors will be preferentially scattered to larger cluster-centric distances, diluting the observed environmental differences \citep{2022ARA&A..60..363N}.
For this part of our work, we only consider galaxies at redshift $z<0.05$.
This redshift cut ensures that the selected sample is complete for galaxies more massive than $\log_{10} \, M_\star/\msun{} \sim 7.5$ and larger than $\log_{10} R_\mathrm{maj}/\mathrm{kpc} \sim 0.3$.
We divide these galaxies into four ranges of $d_\mathrm{cl}$: [0, 2], [2, 10], [10, 30], and [30, 50]~Mpc. In Fig.~\ref{fig:dcl_all}, we plot the size-mass relation for these subsets.
For brevity, we refer to galaxies within 2 Mpc of the nearest cluster center as ``cluster'' galaxies, and at larger separations ($d_\mathrm{cl} > 2$~Mpc) as ``field'' galaxies.

Fig.~\ref{fig:dcl_all} shows that cluster galaxies are systematically smaller than field galaxies at fixed mass, especially for the least massive galaxies in our sample. The differences between different environments are less pronounced at $M_\star > 10^{10.5} \, \msun$. Additionally, we see a hint that galaxies farther from a cluster tend to have larger sizes at almost all masses for which we have significant numbers of galaxies in our sample. As we now demonstrate, the trends in this figure can be understood as the effect of variation in the mix of different galaxy types with mass and environment.

Fig.~\ref{fig:envs_all} shows size-mass relations for subsamples further divided by color (separated by a line drawn on the CMD as in Sect.~\ref{sec:result}) and S\'ersic index ($n > 2.5$ or $n < 2.5$). 
From Fig.~\ref{fig:envs_all} (a) and (b), we find that the size differences between different environments become less prominent when separating the sample by color. Red cluster galaxies are, on average, very slightly smaller than red field galaxies. Blue cluster galaxies are rare and thus harder to compare with the field; only mass bins around $10^8$ to $10^{9.5} \, \msun{}$ contain enough blue cluster galaxies to construct a robust size-mass relation. With this caveat, their sizes appear similar to those of blue galaxies in the field, with a hint that they are slightly larger at $10^{9.5}\,\msun$.
In Fig.~\ref{fig:envs_all} (c) and (d), we see that the offset between cluster and field sizes is more pronounced for low-$n$ galaxies. For high-$n$ galaxies, the difference becomes less prominent with increasing mass.

In the lower row of panels in Fig.~\ref{fig:envs_all}, we further divide red and blue galaxies by S\'ersic index to give four categories: red elliptical ($n>2.5)$, red disk ($n<2.5$), blue elliptical, and blue disk. 
When the sample is separated in this way, we find that none of the subsets show significant differences between the size-mass relations in different environments. The main result of this work is that we find no significant environmental variation within any of these subpopulations, in contrast to the significant differences between their respective average size-mass relations.

Fig.~\ref{fig:envs_morp} emphasizes this result by separating the sample by environment; for each environment, we show the size-mass relation for red elliptical, red disk, blue elliptical, and blue disk galaxies. We find that, in most stellar mass bins, at fixed stellar mass, blue disk galaxies are the largest in size, followed by blue ellipticals, red disks, and finally red ellipticals -- regardless of environment.
In the lower right corner of each panel, we show the fraction of each galaxy type in the corresponding environment. As might be expected, our field galaxy sample is dominated by blue disk galaxies. Notably, we see the fraction of blue galaxies increasing systematically with distance from the nearest cluster even on scales $\gtrsim10$~Mpc. Our cluster galaxy sample is composed mainly of red galaxies (a substantial fraction of which have low $n$). These results reflect the well-known relationship between environment, morphology \citep{dressler_galaxy_1980, poggianti_star_1999, bamford_galaxy_2009, bait_interdependence_2017} and color/star formation rate \citep[e.g.,][]{balogh_bimodal_2004, kauffmann_environmental_2004, weinmann_properties_2006, bamford_galaxy_2009}. This figure emphasizes that our definition of environment, based on the distance to the nearest cluster, yields significantly different mixtures of galaxies in each of our four environment bins.

\section{Discussion}
\label{sec:discussion}

The different mixture of galaxy types in different environments, quantified in Fig.~\ref{fig:envs_morp}, appears to be the dominant factor driving the overall smaller average size for cluster galaxies shown in Fig.~\ref{fig:dcl_all}. The size-mass relation of red galaxies in clusters is no different to that of red galaxies in the field, particularly when low-$n$ and high-$n$ galaxies are considered separately. There is therefore no obvious reason to attribute the size difference to transformative processes that only occur in the cluster environment, such as mass loss \citep{mayer_tidal_2001, boselli_environmental_2006, kazantzidis_efficiency_2010} or tidal effects on the internal structure of galaxies
\citep[tidal heating; ][]{kuchner_effects_2017,fattahi_tidal_2018, lokas_tidal_2020, fielder_all_2024}.

Environmental suppression of star formation \citep[for example due to gas starvation and stripping;][]{gunn_infall_1972, larson_evolution_1980, Font2008MNRAS.389.1619F} may play a role in transforming blue disks to red disks in clusters. However, red disks exist in the field and apparently have a very similar size-mass relation to their counterparts in clusters. This suggests the (substantial) size offset between blue and red disks is determined by an intrinsic (rather than environmental) difference. 
For example, in our sample, although the size ratio between blue and red disks varies with stellar mass, it is on average $\sim1.5$ across environments.
The light-weighted size in a fixed bandpass may be larger in galaxies that are bluer overall, because the light in that bandpass may be dominated by a younger stellar population that is more extended than the total mass. The apparent size of such a galaxy may shrink if star formation is suppressed \citep{cappellari_effect_2013}. This `spurious' evolution of light-weighted size (unrelated to any significant change in the mass distribution) may explain the size difference between red and blue disks in all environments \citep{kelvin2012gama}.  
Alternatively, the smaller sizes of red disks may be due to intrinsically different growth histories -- earlier formation, in more concentrated potentials, followed by correspondingly earlier suppression. The higher background density in clusters implies an earlier average halo collapse time and, hence, potentially older ages at a fixed mass. 

We have only considered the effects of the cluster environment averaged over an aperture of radius 2~Mpc. This `large scale' definition of clusters may be dominated by recently virialized members, which more closely resemble the field. Cluster environmental processes may be more important in the innermost regions of the cluster potential, or for particular combinations of progenitor mass and orbit. Although such effects are interesting probes of the cluster environment itself, our results suggest they are not important for understanding the relationships between mass, color, morphology, and size across the bulk of the observed galaxy population.

Our use of photometric redshifts provides a substantially larger sample than would be possible with spectroscopic redshifts. The systematic trend we see in the mix of subpopulations across environments provides some evidence that our large-scale measure of environment, although imprecise, is still meaningful despite convolution with photo-$z$ errors. We expect these trends to be even clearer with more accurate redshifts. However, photometric redshifts may have significant bias and larger errors for particular subpopulations, such as faint blue galaxies (which may have their luminosity overestimated, as noted above). 
For example, \citet[][]{Ilbert2006A&A...457..841I} and \citet{2022ARA&A..60..363N} show photometric redshift errors are larger for bluer galaxies.

To estimate the blurring of our environmental measure due to photo-$z$ errors, we matched our photometric sample with the DESI DR1 spectroscopic redshift catalog. Size-mass relations obtained using spectroscopic redshifts for galaxies matched to our photo-$z$ sample are noisier and necessarily limited to brighter galaxies, but show the same overall behavior as Fig.~\ref{fig:envs_all} -- the environmental differences within sub-populations are negligible compared to the differences between sub-populations.
Across all matched galaxies, we find a mean redshift difference of $\Delta z\sim0.012$. However, for galaxies with $\log_{10}(M_\star/\msun{}) < 8$ and $g-r > 0.4$, the mean redshift difference is $\Delta z\sim0.045$. It therefore appears that the greatest contribution of photometric redshift errors to our results concerns faint \textit{red} galaxies. These galaxies dominate the red-sequence cluster population but are relatively less common in the field. Consequently, the main systematic effect of photometric redshift error on our sample is to preferentially scatter such galaxies away from clusters. For this reason, we have less confidence in the relations we obtain for the faintest red galaxies.
Future spectroscopic surveys will be better able to test whether inaccuracy and potential biases due to photometric redshifts obscure subtle environmental trends in the size--mass relation within this (or other) populations.

\subsection{Comparison to previous work}

Our analysis is similar to that of \citet[][G24]{ghosh_denser_2024}, who measure the $R_\mathrm{eff}$-stellar mass relation in $360\,\deg^2$ of the HSC-SSP Wide survey area, using photometric redshifts and stellar masses over $0.3<z<0.7$. The lower mass limit of their sample in their lowest redshift bin is $\sim10^{8.5}\,\msun$. They examine the environmental dependence of their relation, based on the spherical top-hat number density within $10$~Mpc (comoving; $\sigma_\mathrm{r,10\,cMpc}$). \citetalias{ghosh_denser_2024} also separate their sample into subpopulations defined by morphology and star formation rate, although with significantly different operational definitions to ours; their morphology metric is the bulge-to-total mass ratio (estimated with a machine learning framework), and their star formation rate is inferred from the SDSS $u-g$ and $r-z$ color space. Unlike our work, \citetalias{ghosh_denser_2024} define each of their subpopulations by a single criterion (e.g.\ they show results for quiescent galaxies and disk-dominated galaxies, but not for galaxies that are both quiescent and disk-dominated). These different sample definitions complicate a direct, quantitative comparison between our work and theirs, because they give rise to systematic differences in the amplitude and slope of the size-mass relation that are comparable to or larger than the differences of interest.

Qualitatively, however, the results of \citetalias{ghosh_denser_2024} are similar to ours overall: they find that the differences between size-mass relations for disk-dominated, bulge-dominated, active, and quiescent galaxies are much larger than those between samples defined by environment within a given subpopulation. Whereas we emphasize the lack of environmental variation, \citetalias{ghosh_denser_2024} instead focus on the small but statistically significant environment trend within each subpopulation. Their conclusion regarding those trends is effectively the opposite of ours. \citetalias{ghosh_denser_2024} find that $R_\mathrm{eff}$ is significantly larger (by up $\sim25\%$) in higher density environments (i.e. at greater $\sigma_\mathrm{r,10\,cMpc})$. This effect is seen across all their subpopulations, and is most obvious at the lower end of their mass and redshift ranges. In contrast, at fixed mass, we find a (weak) systematic trend towards larger $R_\mathrm{maj}$ at lower density. The most direct comparison can be made for our blue and red subpopulations with $n<2.5$, which best sample the low-mass regime. For these subpopulations, the range of average $R_\mathrm{maj}$ over environment at fixed mass is only slightly less than that reported by \citetalias{ghosh_denser_2024}. However, we consider the environmental trends in our data to be weak, because the differences between the environment bins are much smaller than the dispersions within those bins, and also smaller than the likely intrinsic error on $R_\mathrm{maj}$. 

It is possible that the morphology and star formation metrics used in \citetalias{ghosh_denser_2024} are more sensitive to important physical differences that are captured on average by any `blue/red, late-type/early-type' separation. If so, they may be better suited to revealing subtle environmental trends within each subpopulation. For example, our S\'ersic index criterion does not isolate the (rare) population of disk-dominated galaxies with $M_\star \gtrsim10^{11}\,\msun$, which have significantly smaller sizes.\footnote{Unlike the low mass size difference under discussion, it is clear-cut why sizes at the high-mass end of a sample selected explicitly on disk morphology (rather than on a general proxy for the blue cloud, such as star formation rate or S\'ersic index) are smaller than the population average. These are the rare disk galaxies in relatively massive halos that have not experienced a recent major merger despite the relatively high local overdensity implied by their virial mass \citep[e.g.][]{pallero_formation_2025}.} The sense of the trend reported for disk galaxies in \citetalias{ghosh_denser_2024} is nevertheless intriguing, because (as they note) it contradicts the expectation that, at fixed mass, dark matter halo concentrations should be lower in lower density environments \citep[due to later formation and the primary dependence of concentration on mass assembly history e.g.][]{ludlow_mass-concentration-redshift_2016}. Following \citet[][]{mo_formation_1998} and given an environmentally independent distribution of halo angular momentum  \citep[e.g.][]{maccio_concentration_2007}, disks should then be larger in lower density environments. We note that \citet{perez_galaxy_2025}, in a much smaller but carefully constructed sample of cluster and void galaxies, find that early-type galaxies $10^{9} \lesssim M_\star \lesssim 10^{10}\,\msun$ have smaller sizes in voids, whereas late-types are smaller in clusters, particularly at older stellar ages. 

Clearly, further investigation of these discrepancies is warranted. They may simply be the result of differences in size measurement (e.g.\ $R_\mathrm{maj}$ versus $R_\mathrm{eff}$) or sample construction (selection by inferred physical properties such as star formation rate rather than observed properties such as color; single-parameter versus two-parameter subsets). However, if they represent real physical differences, these environmental trends within subpopulations may provide further insight into the relationship between star formation and dark matter halo assembly.

\subsection{Size definition}

As an alternative to  $R_\mathrm{eff}$ and other empirical measures of galaxy size,
\citet{trujillo_physically_2020} have proposed a physically motivated definition based on the maximum extent of gas capable of supporting \textit{in situ} star formation. They define $R_1$, the radius of a stellar mass surface density contour at $1\,\mathrm{\msun{}\,pc^{-2}}$, as an observable proxy for this size \citep[see e.g.][]{chamba_historical_2020, chamba_edges_2022, chamba_impact_2024, buitrago_strong_2024, golini_lights_2025}. They have shown that scaling relations for $R_1$ have lower scatter and are closer to linear across a broader range of mass, compared to relations based on $R_\mathrm{eff}$. $R_1$ is likely to be less sensitive to `higher order' processes, such as bulge formation, that lead to differences in stellar concentration between galaxies with similar present-day viral masses and assembly histories. $R_1$  may therefore be a more robust basis for comparisons between different galaxy types and between observed and simulated galaxies, with the aim of understanding the most important factors controlling size at fixed stellar mass \citep{arjona-galvez_physically_2025}. At present, however, the surface brightness at $R_1$ is close to or below the limit for the majority of galaxies in the largest imaging surveys. Moreover, the greater complexity reflected in scaling relations for $R_\mathrm{eff}$ may be particularly important for constraining the relationship between (for example) structural evolution and the shutdown of star formation. From a theoretical perspective, we believe it is important for simulations to demonstrate agreement with both $R_\mathrm{eff}$-like and $R_1$-like measures of size.

\section{Summary and conclusions}
\label{sec:summary}

In this paper, we study the size-mass relation of galaxies at $z<0.3$ using the photometric catalog from the DESI Legacy Imaging Survey DR10. We further select a subsample of galaxies with $z<0.05$ and investigate the size changes in different environments, defined by distance to the nearest cluster in the catalog of WH2024. 
Our main results are as follows:

\begin{itemize}
    \item We use maximum likelihood estimates of average size for a $V/V_\mathrm{max}$ weighted sample to obtain the relation between semi-major axis size and luminosity. The results are in good agreement with previous studies    (Fig.~\ref{fig:Mr_size}).
    \item Using galaxies in both the GSWLC catalog \citep{salim_dust_2018} and WH2024, we obtain the size-mass relation for the mass range $10^7 \leq M_\star \leq 10^{12.5} \, \msun$. The logarithmic slope of the overall size-mass relation is $\sim0.3$, consistent with previous studies (Fig.~\ref{fig:size_mass}).
    \item We examine the consistency between our size-mass relation and samples of individual galaxy sizes in the literature, from $\sim10^5$ to $10^{12.5} \, M_\odot$ (Figs.~\ref{fig:MVsize} and \ref{fig:sizemass_overlay}). In the dwarf galaxy regime ($<10^{9}\,\msun$), we see hints of a steepening relation consistent with the extrapolation of relations for small samples of nearby dwarf galaxies (dominated by satellites); however, different samples show an apparently increasing scatter towards lower mass. This may be due to sample selection effects as well as physical differences.
    \item For galaxies at $z< 0.05$, we computed the distance to the nearest cluster center. We found that galaxies closer to clusters, in general, have smaller sizes than field galaxies. (Fig.~\ref{fig:dcl_all}).
    We separate galaxies into four sub-populations with red/blue colors and high/low S\'ersic index. Figs.~\ref{fig:envs_all} and \ref{fig:envs_morp} show that although the relative fraction, of the subpopulations clearly vary with the distance to the nearest cluster, the size-mass relation of each subpopulation does not. This is the main result of this paper.
\end{itemize}

Given the complexity of the galaxy population and potential environmental effects on size, our findings are remarkably simple: they are consistent with environmental variations in the size-mass relation that are due only to variation in the mixture of the color-morphology subpopulations with cosmic density. In turn, this suggests that the assembly history of a galaxy has more influence on its size than any direct effects of its cosmic environment, such as gas starvation and tidal heating/stripping. 

Our results are qualitatively consistent with an extensive literature that favors internal evolution (ultimately determined by the present-day mass and assembly history of the host halo) as the primary driver of bimodality in galaxy structure and star formation rate, with only a minor role for environmental effects \citep{schawinski_green_2014, lokas_formation_2022}. In detail, the processes driving this evolution remain unclear, even in simulations \citep[e.g.][]{walters_structural_2021, walters_quenching_2022,lokas_formation_2022, somerville_relationship_2025}. The results we show in Figs.~\ref{fig:envs_all}-\ref{fig:envs_morp} and tabulate in Appendix~\ref{sec:tables} are based on straightforward definitions of subpopulations selected using color and S\'ersic index. Selections based on `apparent' observable quantities like these, for large samples, are useful because they can be compared in a direct and robust way to model predictions. Such comparisons require forward-modelling of fundamental simulated quantities, like star formation rate and stellar mass density, into nontrivial observables. This is often challenging. However, the forward-modelling approach is still highly complementary to the alternative of comparing physical properties inferred from the data to `raw' simulated quantities \citep[as, for example, we argue in][]{liao_colour_2022}. Ultimately, agreement between the two approaches would give confidence that both the data and the simulations are sufficiently well understood.

Further study of the $R_\mathrm{maj}$-mass relation will be particularly interesting in the low-mass field. Fig.~\ref{fig:sizemass_overlay} shows that DESI-LS begins to bridge the gap between wide-field studies of the overall galaxy population and samples of dwarf galaxies in the nearby Universe -- so far, these are mostly satellites of $L^\star$ galaxies. We see hits of a steepening slope in the size-mass relation around $M_\star\sim10^{8}$, which would match the steep slope of the small-scale relation. This is broadly consistent with previous work at this mass scale. However, the amplitude and scatter of the relation at this scale are poorly understood in theory, and not obviously consistent between different observational datasets. At lower masses, centrifugal support may not be the critical factor controlling galaxy size, and the apparent diversity of dwarf galaxy morphologies and star formation histories remains something of a mystery, even for field galaxies. Although new data from Euclid and Rubin/LSST will almost certainly lead to substantial progress on these questions, we believe our results demonstrate that further advances are already possible with imaging from DESI-LS. Larger spectroscopic samples from the DESI Bright Galaxy Survey will be particularly helpful to address uncertainties associated with the use of photometric redshifts, particularly for the faintest galaxies in our sample.

\section*{Data availability}

The size-magnitude and the size-mass relation shown in Fig.~\ref{fig:Mr_size} and \ref{fig:size_mass} are provided at \url{https://osf.io/yhzfv/} in the format of FITS files. We also provide the size-mass relation for galaxies in the cluster of each subset as shown in Fig.~\ref{fig:envs_morp}. The files are also available at the CDS via anonymous ftp to \url{cdsarc.u-strasbg.fr} (130.79.128.5) or via \url{http://cdsweb.u-strasbg.fr/cgi-bin/qcat?J/A+A/}.

\begin{acknowledgements}

The authors thank Mireia Montes and Pablo Fosalba for useful discussions, and the anonymous referee for their insightful and constructive feedback. LWL acknowledges support from the Arrakihs mission through grant number PID2022-138896NB-C52 /MICIN/AEI/10.13039/501100011033/FEDER, UE; the Spanish Ministerio de Ciencia, Innovaci\'on y Universidades, project PID2022-138896NB; the European Research Executive Agency HORIZON-MSCA-2021-SE-01 Research and Innovation programme under the Marie Skłodowska-Curie grant agreement number 101086388 (LACEGAL) and the programme Unidad de Excelencia Mar\'{\i}a de Maeztu, project CEX2020-001058-M.
APC is supported by a Taiwan Ministry of Education (MoE) Yushan Fellowship, MOE-113-YSFMS-0002-001-P2, and acknowledges Taiwan National Science and Technology Council (NSTC) grants 112-2112-M-007-017 and 112-2112-M-007-009. This work used high-performance computing facilities operated by the Center for Informatics and Computation in Astronomy (CICA) at National Tsing Hua University. This equipment was funded by MoE, NSTC, and National Tsing Hua University.

The Legacy Surveys consist of three individual and complementary projects: the Dark Energy Camera Legacy Survey (DECaLS; Proposal ID \#2014B-0404; PIs: David Schlegel and Arjun Dey), the Beijing-Arizona Sky Survey (BASS; NOAO Prop. ID \#2015A-0801; PIs: Zhou Xu and Xiaohui Fan), and the Mayall $z$-band Legacy Survey (MzLS; Prop. ID \#2016A-0453; PI: Arjun Dey). DECaLS, BASS and MzLS together include data obtained, respectively, at the Blanco telescope, Cerro Tololo Inter-American Observatory, NSF’s NOIRLab; the Bok telescope, Steward Observatory, University of Arizona; and the Mayall telescope, Kitt Peak National Observatory, NOIRLab. Pipeline processing and analyses of the data were supported by NOIRLab and the Lawrence Berkeley National Laboratory (LBNL). The Legacy Surveys project is honored to be permitted to conduct astronomical research on Iolkam Du’ag (Kitt Peak), a mountain with particular significance to the Tohono O’odham Nation.

NOIRLab is operated by the Association of Universities for Research in Astronomy (AURA) under a cooperative agreement with the National Science Foundation. LBNL is managed by the Regents of the University of California under contract to the U.S. Department of Energy.

This project used data obtained with the Dark Energy Camera (DECam), which was constructed by the Dark Energy Survey (DES) collaboration. Funding for the DES Projects has been provided by the U.S. Department of Energy, the U.S. National Science Foundation, the Ministry of Science and Education of Spain, the Science and Technology Facilities Council of the United Kingdom, the Higher Education Funding Council for England, the National Center for Supercomputing Applications at the University of Illinois at Urbana-Champaign, the Kavli Institute of Cosmological Physics at the University of Chicago, Center for Cosmology and Astro-Particle Physics at the Ohio State University, the Mitchell Institute for Fundamental Physics and Astronomy at Texas A\&M University, Financiadora de Estudos e Projetos, Fundacao Carlos Chagas Filho de Amparo, Financiadora de Estudos e Projetos, Fundacao Carlos Chagas Filho de Amparo a Pesquisa do Estado do Rio de Janeiro, Conselho Nacional de Desenvolvimento Cientifico e Tecnologico and the Ministerio da Ciencia, Tecnologia e Inovacao, the Deutsche Forschungsgemeinschaft and the Collaborating Institutions in the Dark Energy Survey. The Collaborating Institutions are Argonne National Laboratory, the University of California at Santa Cruz, the University of Cambridge, Centro de Investigaciones Energeticas, Medioambientales y Tecnologicas-Madrid, the University of Chicago, University College London, the DES-Brazil Consortium, the University of Edinburgh, the Eidgenossische Technische Hochschule (ETH) Zurich, Fermi National Accelerator Laboratory, the University of Illinois at Urbana-Champaign, the Institut de Ciencies de l’Espai (IEEC/CSIC), the Institut de Fisica d’Altes Energies, Lawrence Berkeley National Laboratory, the Ludwig Maximilians Universitat Munchen and the associated Excellence Cluster Universe, the University of Michigan, NSF’s NOIRLab, the University of Nottingham, the Ohio State University, the University of Pennsylvania, the University of Portsmouth, SLAC National Accelerator Laboratory, Stanford University, the University of Sussex, and Texas A\&M University.

BASS is a key project of the Telescope Access Program (TAP), which has been funded by the National Astronomical Observatories of China, the Chinese Academy of Sciences (the Strategic Priority Research Program “The Emergence of Cosmological Structures” Grant \#XDB09000000), and the Special Fund for Astronomy from the Ministry of Finance. The BASS is also supported by the External Cooperation Program of Chinese Academy of Sciences (Grant \#114A11KYSB20160057), and Chinese National Natural Science Foundation (Grant \#12120101003, \#11433005).

The Legacy Survey team makes use of data products from the Near-Earth Object Wide-field Infrared Survey Explorer (NEOWISE), which is a project of the Jet Propulsion Laboratory/California Institute of Technology. NEOWISE is funded by the National Aeronautics and Space Administration.

The Legacy Surveys imaging of the DESI footprint is supported by the Director, Office of Science, Office of High Energy Physics of the U.S. Department of Energy under Contract No. DE-AC02-05CH1123, by the National Energy Research Scientific Computing Center, a DOE Office of Science User Facility under the same contract; and by the U.S. National Science Foundation, Division of Astronomical Sciences under Contract No. AST-0950945 to NOAO.
\end{acknowledgements}

%
   \bibliographystyle{aa} 
   \bibliography{references} 
%
\begin{appendix}

\section{k-correction}
\label{appendix:kcorrection}
\begin{figure*}[htbp]
    \centering
    \includegraphics[width=0.88\textwidth, trim=0 20 0 0]{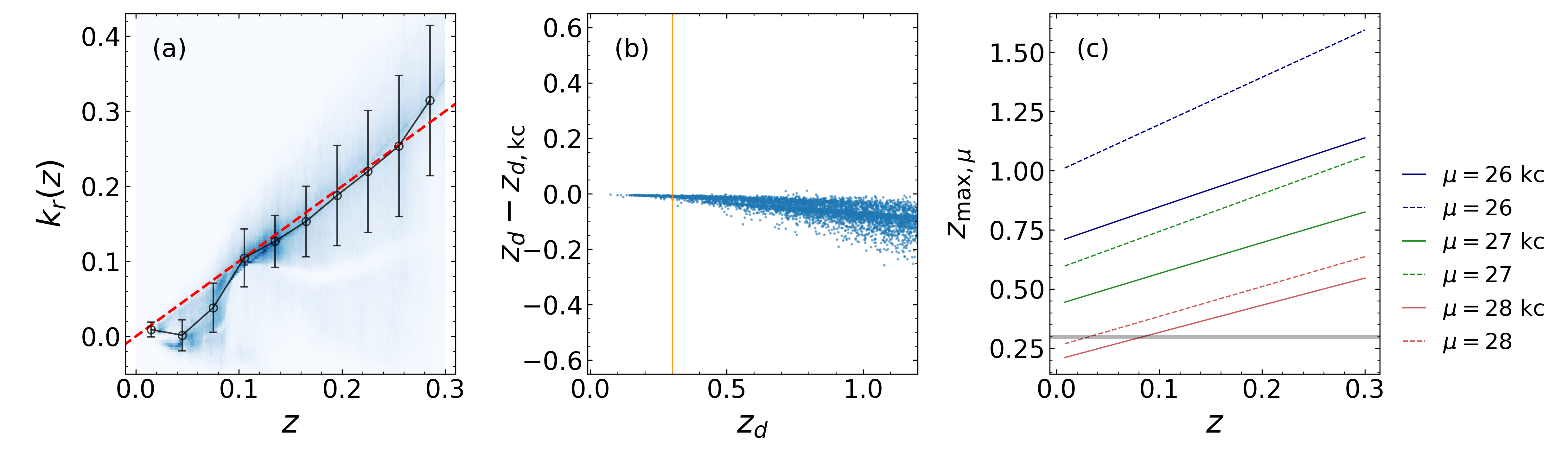}
    \caption{(a) The $k$-correction in the $r$ band as a function of redshift. The blue scale shows the histogram of the distribution, while the black dots indicate the mean values in redshift bins. The error bars represent the standard deviation within each bin. The red dashed line labels the $k_r(z) \sim z$ relation.
    (b) Redshift from the maximum luminosity distance versus the difference between values without and with the $k$-correction. The vertical orange line indicates $z = 0.3$.
    (c) Maximum redshift set by the surface brightness limit ($z_\mu$) versus redshift. Solid lines show $z_\mu$ obtained with the $k$-correction, assuming $k(z)\approx z$ for illustration, while dashed lines show $z_\mu$ computed without $k$-correction. Different colors indicate results for galaxies with different surface brightnesses. For this test, we set $\mu_\mathrm{lim}=29$. The horizontal gray line indicates $z=0.3$, the maximum redshift limit we impose on our sample.
    }
    \label{fig:kcplot}
\end{figure*}

In this appendix, we explain why we neglect the $k$-correction when computing $1/V_\mathrm{max}$ weights in Eqs.~\ref{eq:dL} and \ref{eq:zmu}. Fig.~\ref{fig:kcplot} (a) shows the $k$-correction (computed using the Python package \texttt{kcorrect} from \citealt{blanton_k_2007}) as a function of redshift. The value increases with redshift as expected. Empirically, the $k$-correction scales approximately as $k(z) \sim z$ \citep[see also][]{Hogg2002astro.ph.10394H, omill_photometric_2011}.

The maximum redshift at which a given galaxy enters our sample is determined by the smallest of $z_d$, $z_\mu$, $z_A$, and the upper redshift limit of 0.3. Among these limiting redshifts, $z_d$ and $z_\mu$ are affected by the $k$-correction. The luminosity distance including the $k$-correction is given by
\begin{equation}
    d_{L,\mathrm{max,kc}}=10^{-0.2(M-m_\mathrm{lim}+k(z)-5)},
\end{equation}
where $M$ is the absolute magnitude of a galaxy, $m_\mathrm{lim}$ is the limiting apparent magnitude, and $k(z)$ is the $k$-correction.

To test whether the $k$-correction introduces significant differences in our estimate of $V_\mathrm{max}$, we compute the luminosity distance with and without the $k$-correction (we refer to these values as $d_{L,\mathrm{max}}$ and $d_{L,\mathrm{max,kc}}$ respectively). From these distances we obtain the corresponding limiting redshifts, $z_d$ and $z_{d,\mathrm{kc}}$. Fig.~\ref{fig:kcplot} (b) shows $z_d - z_{d,\mathrm{kc}}$ versus $z_d$. The differences become significant for galaxies with $z_d > 0.3$, but are negligible for $z_d < 0.3$ (the mean difference is $\sim 0.007$~mag). Since we impose a maximum redshift of $z=0.3$, galaxies with $z_d > 0.3$ do not affect our results. We therefore do not include a $k$-correction term in our calculation of $V_\mathrm{max}$.

The calculation of $V_\mathrm{max}$ accounting for the $k$-correction is more complicated when the redshift limit is set by the surface brightness limit, $z_\mu$, since both the observed surface brightness and the limited surface brightness require $k$-corrections whose values depend on redshift. We demonstrate this from the surface brightness after applying the $k$-correction is
\begin{equation}
    \mu_{\mathrm{kc}} = \mu+10 \log_{10} (1+z) + k(z),
\end{equation}
where $\mu_{\mathrm{kc}}$ is the $k$-corrected surface brightness, and $\mu$ and $z$ are the observed surface brightness and redshift of the galaxy.
We can then compute the difference between the observed $k$-corrected $\mu$ and the $k$-corrected $\mu_\mathrm{lim}$:
\begin{equation}
    \mu_\mathrm{obs}-\mu_\mathrm{lim} = 10 \log_{10}\frac{1+z_\mathrm{lim}}{1+z} + k(z_\mathrm{lim})-k(z)
\end{equation}
This can be solved numerically for $z_\mathrm{lim}$. To illustrate the redshift dependence of $z_\mu$ for a galaxy of given surface brightness, we substitute $k(z_\mathrm{lim}) \approx z_\mathrm{lim}$ and $k(z) \approx z$, as shown in Fig.~\ref{fig:kcplot} (a), and compute the corresponding $z_\mu$ for a given $z$ and $\mu$.
Fig.~\ref{fig:kcplot} (c) shows the resulting $z_\mu$. 
The differences in $z_\mu$ are more prominent for higher surface brightness galaxies, but most galaxies with high surface brightness have limiting redshifts far beyond the upper limit we impose at $z=0.3$ regardless of the $k$-correction. The $k$-correction therefore does not affect the final determination of $z_\mathrm{max}$ for the majority of galaxies. The same is true for lower surface brightness galaxies at $z\gtrsim0.1$. Therefore, only nearby galaxies ($z \lesssim 0.1$) with $\mu \gtrsim 27.5$ are affected. 

\section{Size and magnitude of LS and SDSS}
\label{appendix:ls_sdss}

\begin{figure}[htbp]
    \centering
    \includegraphics[width=0.45\textwidth, trim=0 20 0 0]{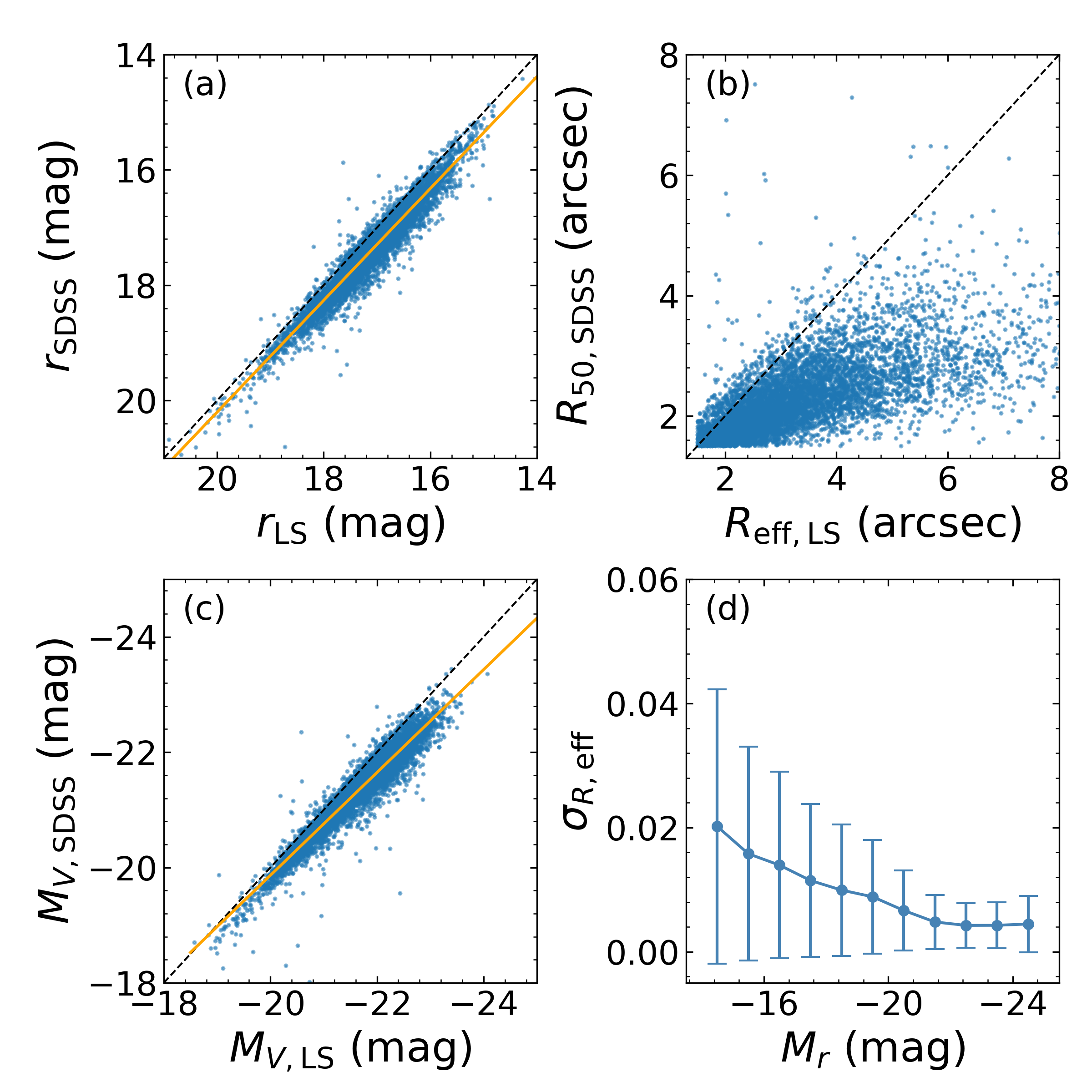}
    \caption{(a) Apparent $r$-band magnitude from LS and $r-$band magnitude obtained from SDSS of randomly selected galaxies. 
    (b) Half-light radius from \texttt{TRACTOR} catalog of the LS survey and the $r-$band Petrosian half-light radius from SDSS of the randomly selected galaxies.
    (c) $M_V$ converted using $r_\mathrm{LS}$ and $r_\mathrm{SDSS}$.
    The black dashed lines in both panels are one-to-one relations. The orange line is the fitted relation of $r$-band magnitudes.
    (d) The mean error of the modeled size in LS as a function of $M_r$. The error bars represent the standard deviation of the error within each magnitude bin.} 
    \label{fig:ls_sdss}
\end{figure}

\begin{figure}
    \centering
    \includegraphics[width=0.45\textwidth, trim=0 15 0 0]{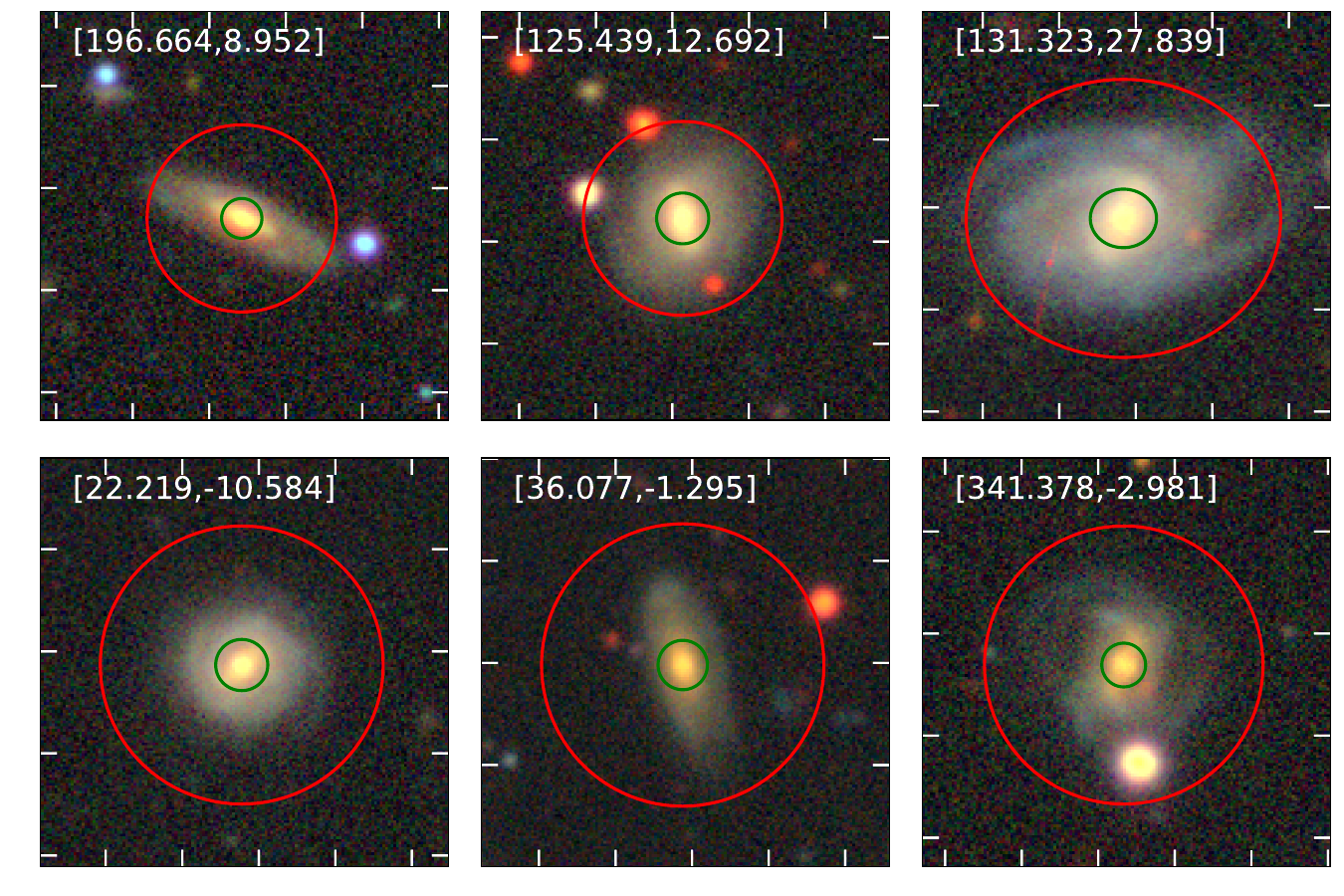}
    \caption{Example galaxies for which the sizes measured in LS are much larger than the Petrosian sizes from SDSS. The RA and Dec of the galaxies are shown in the upper-left corner of each panel. The green circle indicates the SDSS Petrosian size, while the red circle marks the size fitted in LS.}
    \label{fig:ex_diffr}
\end{figure}

\begin{figure}
    \centering
    \includegraphics[width=0.98\linewidth, trim=0 20 0 0]{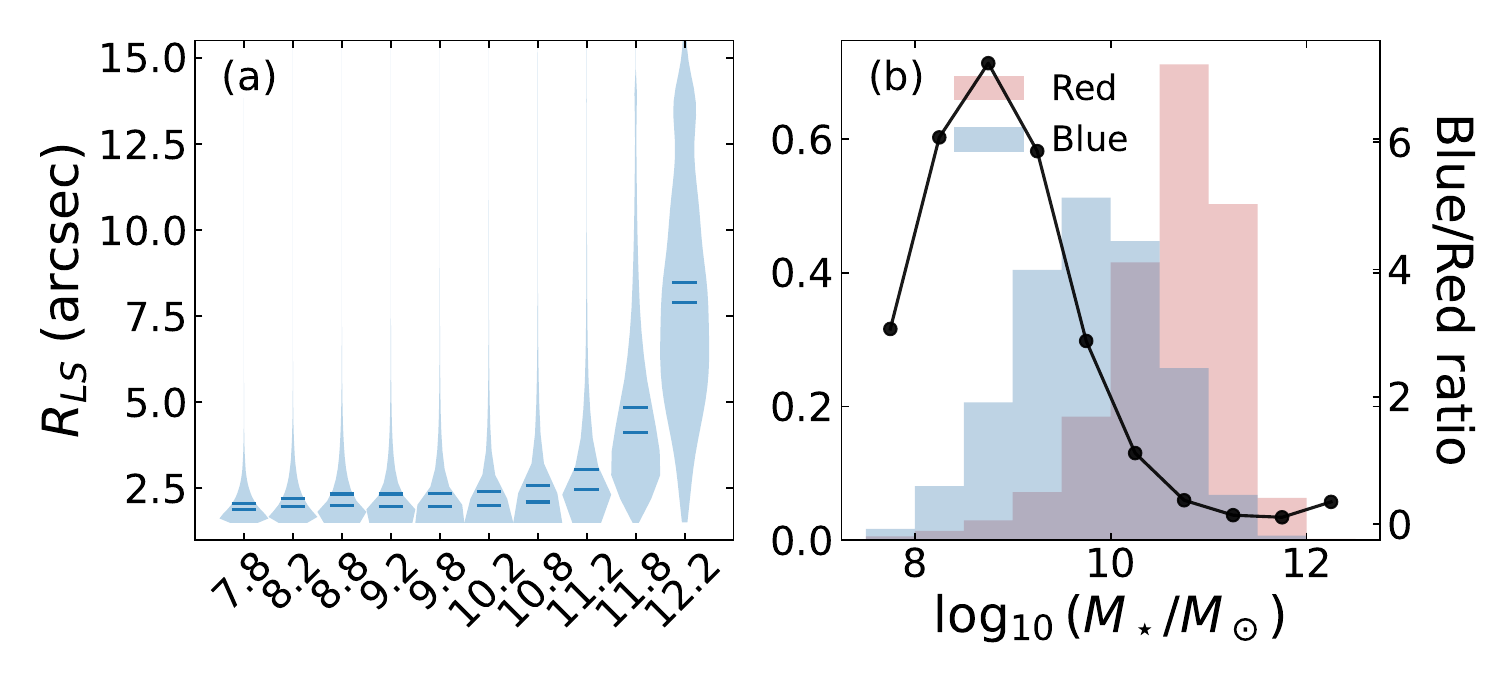}
    \caption{The upper panel shows the distribution of size in arcseconds for each stellar mass bin. The horizontal bars show the 25th and 75th percentiles of the size distribution. The histogram in the bottom panel shows the stellar mass distribution for red and blue galaxies. The black lines show the ratio between blue and red galaxies in each mass bin.}
    \label{fig:diffr_violin}
\end{figure}

In Figs.~\ref{fig:MVsize} and \ref{fig:sizemass_overlay}, we compare our size--mass relation to literature size measurements for individual dwarf galaxies. For these comparisons, we convert our magnitudes to the $V$-band using the LS $g$ and $r$ magnitudes. However, the relation we use (that of \citealt{carlsten_structures_2021}) was fit to the SDSS $g$ and $r$ bands. There are small differences between the two systems. Due to the different survey footprints and greater depth of LS, we do not have SDSS photometry for many of our \textit{pz} sample galaxies. Relations are available to convert SDSS (or PanSTARRS) photometry to the LS  system, but not vice versa (i.e., the available relations are limited to the magnitude range of SDSS) and they require the SDSS (or PanSTARRS) $g-i$ color. Since our application is very simple, we decided not to try to adapt these relations, or to derive a new $V$-band conversion from the LS photometry. Instead, we simply treat the LS bands as equivalent to the SDSS bands for the purpose of the comparison in Figs.~\ref{fig:MVsize} and \ref{fig:sizemass_overlay}.

To explore the impact of this choice, we randomly selected 10,000 SDSS galaxies and matched them to LS. In total, this matched test sample contains 8,031 galaxies.
Fig.~\ref{fig:ls_sdss}a compares the $r$-band magnitude from LS ($r_\mathrm{LS}$) with the $r$-band magnitude from SDSS ($r_\mathrm{SDSS}$). At a given $r_\mathrm{LS}$, $r_\mathrm{SDSS}$ is approximately 0.2–0.3 magnitudes fainter. A linear fit between $r_\mathrm{LS}$ and $r_\mathrm{SDSS}$ yields the relation: $r_\mathrm{SDSS} = 0.97 \times (r_\mathrm{LS}-20) + 20.20$.

Fig.~\ref{fig:ls_sdss} (b) shows $R_\mathrm{eff}$ from LS and Petrosian $R_{50}$ from the SDSS catalog. This figure demonstrates that the sizes measured in SDSS differ from those in LS.
Previous studies have shown that Petrosian magnitudes systematically underestimate sizes for galaxies with high Sérsic indices \citep{bernardi_systematic_2014}, and detailed theoretical studies on this issue have been provided by \citet{graham_concise_2005} and \citet{graham_total_2005}.
In addition to the intrinsic differences between model-based and Petrosian radii, errors in radius fitting may also contribute to these discrepancies. Fig.~\ref{fig:ls_sdss} (d) shows the mean error of the modeled radius in bins of $M_r$ for LS galaxies, indicating that the errors become larger for faint galaxies.

We show in Fig.~\ref{fig:ex_diffr} examples of galaxies with large size differences between LS and SDSS. Most of these are disk galaxies. The Petrosian radius from SDSS tends to cover only the bulge region, while the sizes fitted from LS tend to cover the entire disk structure. The LS sizes appear to overestimate the true galaxy size in these cases. To test whether this affects the size–mass relation, we plot a violin diagram in stellar mass bins in Fig.~\ref{fig:diffr_violin} (a). 
We find that in most stellar mass bins, the galaxy sizes are smaller than $\sim$4 arcsec; we regard this as a `safe' regime in which sizes from LS and SDSS are in reasonable agreement (see Fig.~\ref{fig:ls_sdss} (b)). 
Only galaxies with $\log_{10}(M_\star/M_\odot) > 11$ appear to be affected by severe overestimates. However, Fig.~\ref{fig:diffr_violin} (b) shows that the fraction of blue galaxies (predominantly disks) is low in those high-mass bins. Therefore, we conclude that our results are not significantly affected by these overestimated sizes.

Fig.~\ref{fig:ls_sdss} (c) shows the $M_V$ values derived from Equation~\ref{eq:vband} using $g$ and $r$ from LS and SDSS. At a fixed $M_{V, \mathrm{SDSS}}$, the $V$-band magnitude derived from LS filters is brighter, with the differences becoming more pronounced for brighter galaxies. We also fit a relation between $M_{V,\mathrm{SDSS}}$ and $M_{V,\mathrm{LS}}$ as follows: $M_{V,\, \mathrm{SDSS}} = 0.89 \times (M_{V, \,\mathrm{LS}}+22) - 21.65$.
This very simple relation only captures the average difference between $V$-band magnitudes estimated with $g$ and $r$ filters from the two surveys. The difference is small for galaxies fainter than $M_V=-20$, which is the regime of interest in Figs.~\ref{fig:MVsize} and \ref{fig:sizemass_overlay}.

\section{Dispersion in size due to morphology at fixed magnitude}
\label{appendix:sigma_n}

\begin{figure*}[htbp]
    \sidecaption
    \includegraphics[width=12cm]{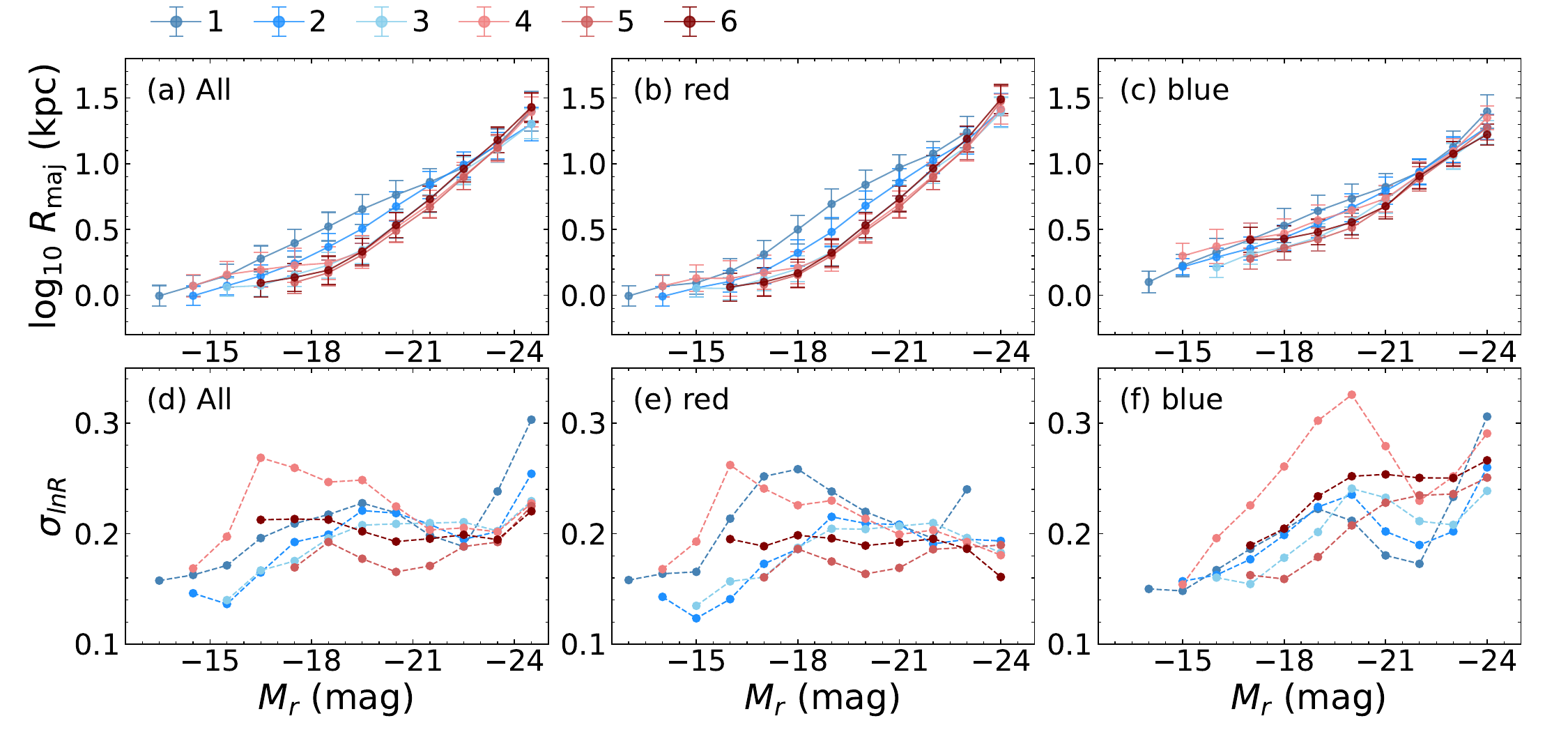}
    \caption{$M_r$ vs. $R_\mathrm{maj}$ relation (top panels) and scatter of size distributions (lower panels) for galaxies with different S\'ersic indices. From left to right are red, blue, and all the galaxies. The red and blue galaxies are separated following Fig.~\ref{fig:cmd}. The error bars are the error in the likelihood fitting. The S\'ersic indices of each group are [$0.5-1.5$, $1.5-2.5$, $2.5-3.5$, $3.5-4.5$, $4.5-5.5$, $5.5-6.5$].}
    \label{fig:size_lumi_n}
\end{figure*}

The variation of the dispersion in size with absolute magnitude has been used to constrain galaxy evolution models, and has also been shown to be related to galaxy morphology \citep{shen_size_2003, graham_inclination-_2008, avila-reese_baryonic_2008, saintonge_disk_2011, bernardi_systematic_2014, hon_size-mass_2023}.
For example, \citet{bernardi_systematic_2014}, discuss the scatter in the size-luminosity relation for different types of galaxy. 
We obtain similar results when separating galaxies by S\'ersic index. 

Fig.~\ref{fig:size_lumi_n} shows the size-mass relation, and the dispersion around that relation, in bins of different Sérsic indices for all, red, and blue galaxies. For both blue and red galaxies, those with a lower Sérsic index tend to have larger sizes at a fixed $M_r$, particularly in the range $-18 > M_r > -21$. This suggests that the presence of a disk influences the scatter in the size-magnitude relation,  and hence that variation in scatter with $M_r$ should reflect variation of the morphological mix with $M_r$.

For blue galaxies, size dispersion increases with magnitude, largely independent of Sérsic index but with an interesting minimum for later ($n<3$) types around $M_{r}\approx-21.5$. The prominence of this downturn correlates with Sérsic index up to $n\sim4$. As we discuss in Sect.~\ref{sec:Mrsize}, the overall increase may reflect the increasing contribution of central bulges $R_\mathrm{maj}$ in brighter galaxies. We speculate the downturn may then be the result of the bluest, most massive disks being those with the lowest bulge fractions.

The dispersions of red galaxies are approximately constant and approximately independent of Sérsic brighter than $M_{r}\approx-19$. Curiously, there is no sign of a downturn for later types at brighter magnitudes. At $M_{r}\gtrsim-19$, red galaxies with $2<n<3$ have comparable dispersion to blue galaxies, but those with $n=1$ show a peak in dispersion around $M_{r}\approx-17.5$. This is difficult to explain; it may point to a more diverse population of non-star-forming low-surface-brightness galaxies in this magnitude range.

Finally, we note that the dispersion curves for $n=4$ seem to deviate strongly from the overall trend with Sérsic index for both the blue and red subsets. Loosely, the dispersion for the $n=4$ subset appears to increase in proportion to the spread of dispersion among the other subsets. The reason for this is also unclear; it may be a systematic related to the fitting procedure in \texttt{Tractor}. Further investigation of these trends in dispersion is beyond the scope of this work, but the difficulty of interpreting them suggests that such investigation would be worthwhile.

\section{Table and data}
\label{sec:tables}
\begin{table*}[htbp]
\small
\caption{Parameters of fits to the size-magnitude relation (equation~\ref{eq:sizemag}) and size-mass relation (equation~\ref{eq:sizemasslog}).\label{tab:size_Mr}}
\centering
\begin{tabular}{lcccccc}
\hline\hline
& \multicolumn{2}{c}{$R_\mathrm{eff}-M_r$}& \multicolumn{2}{c}{$R_\mathrm{eff,major}-M_r$} & \multicolumn{2}{c}{$R_\mathrm{eff,major}-M_r$} \\
& \multicolumn{2}{c}{\sz} & \multicolumn{2}{c}{\sz} & \multicolumn{2}{c}{\pz} \\
\cline{2-7}
& $a$ & $b$ & $a$ & $b$ & $a$ & $b$ \\
\hline
All & $-0.172\pm0.012$ & $-2.922\pm0.249$ 
& $-0.163\pm0.014$ & $-2.612\pm0.300$ 
& $-0.121\pm0.007$ & $-1.661\pm0.137$ 
\\
Red & $-0.175\pm0.011$ & $-3.009\pm0.238$ 
& $-0.168\pm0.014$ & $-2.747\pm0.295$ 
& $-0.126\pm0.008$ & $-1.855\pm0.142$ 
\\
Blue & $-0.172\pm0.013$ & $-2.896\pm0.263$ 
& $-0.164\pm0.014$ & $-2.609\pm0.309$ 
& $-0.121\pm0.007$ & $-1.653\pm0.137$ 
\\
$n>2.5$ 
& $-0.191\pm0.011$ & $-3.382\pm0.243$ 
& $-0.187\pm0.013$ & $-3.198\pm0.274$ 
& $-0.136\pm0.007$ & $-1.942\pm0.128$ 
\\
$n<2.5$ 
& $-0.155\pm0.012$ & $-2.569\pm0.252$ 
& $-0.147\pm0.014$ & $-2.283\pm0.307$ 
& $-0.108\pm0.007$ & $-1.404\pm0.120$ 
\\
\hline\hline
& \multicolumn{6}{c}{$R_\mathrm{eff, major}-M_\star$} \\
& \multicolumn{2}{c}{\sz} & \multicolumn{2}{c}{$M_\star \geq 10^{10.5} \, M_\odot$} & \multicolumn{2}{c}{$M_\star \leq 10^{10.5} \, M_\odot$} \\
\cline{2-7}
& $a$ & $b$ & $a$ & $b$ & $a$ & $b$ \\
\hline
All
& $0.313\pm0.023$ & $-2.532\pm0.247$ 
& $0.425\pm0.070$ & $-3.864\pm0.827$ 
& $0.275\pm0.059$ & $-2.180\pm0.539$ 
\\
Red
& $0.362\pm0.034$ & $-3.111\pm0.380$ 
& $0.444\pm0.072$ & $-4.088\pm0.845$ 
& $0.249\pm0.118$ & $-2.011\pm1.140$ 
\\
Blue
& $0.298\pm0.028$ & $-2.362\pm0.291$ 
& $0.348\pm0.100$ & $-2.941\pm1.131$ 
& $0.311\pm0.054$ & $-2.471\pm0.502$ 
\\
$n>2.5$
& $0.425\pm0.036$ & $-3.883\pm0.400$ 
& $0.506\pm0.063$ & $-4.843\pm0.740$ 
& $0.268\pm0.118$ & $-2.329\pm1.169$ 
\\
$n<2.5$
& $0.287\pm0.028$ & $-2.268\pm0.286$ 
& $0.256\pm0.101$ & $-1.917\pm1.143$ 
& $0.307\pm0.057$ & $-2.444\pm0.520$ 
\\
\hline\hline
& \multicolumn{6}{c}{$R_\mathrm{eff, major}-M_\star$} \\
& \multicolumn{2}{c}{\pz} & \multicolumn{2}{c}{$M_\star \geq 10^{10.5} \, M_\odot$} & \multicolumn{2}{c}{$M_\star \leq 10^{10.5} \, M_\odot$} \\
\cline{2-7}
& $a$ & $b$ & $a$ & $b$ & $a$ & $b$ \\
\hline
All
& $0.248\pm0.016$ & $-1.821\pm0.153$ 
& $0.399\pm0.081$ & $-3.556\pm0.936$ 
& $0.213\pm0.032$ & $-1.538\pm0.262$ 
\\
Red
& $0.285\pm0.019$ & $-2.280\pm0.191$ 
& $0.486\pm0.069$ & $-4.581\pm0.794$ 
& $0.155\pm0.039$ & $-1.177\pm0.343$ 
\\
Blue
& $0.236\pm0.019$ & $-1.703\pm0.170$ 
& $0.207\pm0.103$ & $-1.376\pm1.159$ 
& $0.236\pm0.030$ & $-1.710\pm0.248$ 
\\
$n>2.5$
& $0.255\pm0.016$ & $-1.933\pm0.150$ 
& $0.435\pm0.082$ & $-3.974\pm0.946$ 
& $0.180\pm0.032$ & $-1.333\pm0.262$ 
\\
$n<2.5$
& $0.222\pm0.019$ & $-1.520\pm0.180$ 
& $0.297\pm0.073$ & $-2.383\pm0.836$ 
& $0.209\pm0.035$ & $-1.412\pm0.306$ 
\\
\hline
\end{tabular}
\end{table*}

We fit the size-magnitude and size-mass relations for the \sz{} and \pz{} samples, separating them into early-type and late-type galaxies based on the Sérsic index and into `red' and `blue' categories based on the CMD (Table~\ref{tab:size_Mr}).

\end{appendix}

\end{document}